%% file: ms.tex
\documentclass[useAMS,usenatbib]{mn2e}

\usepackage{graphicx}
\usepackage{epsfig}
\usepackage{natbib}
\input macros.tex
\begin{document}

\title[The Role of Radiation Pressure in Early Dwarf Galaxies]{The
  Birth of a Galaxy. II. The Role of Radiation Pressure}

\author[J. H. Wise et al.]{John H. Wise$^1$\thanks{e-mail:
    jwise@physics.gatech.edu}, Tom Abel$^2$, Matthew J. Turk$^3$, Michael
  L. Norman$^4$ and \newauthor
  Britton D. Smith$^5$\\
  $^{1}$ Center for Relativistic Astrophysics,
  Georgia Institute of Technology, 837 State Street, Atlanta, GA
  30332, USA\\
  $^{2}$ Kavli Institute for Particle Astrophysics and Cosmology,
  Stanford University, Menlo Park, CA 94025, USA\\
  $^{3}$ Department of Astronomy, Columbia University, 538 West 120th
  Street, New York, NY 10027, USA\\
  $^{4}$ Center for Astrophysics and Space Sciences, University
  of California at San Diego, La Jolla, CA 92093, USA\\ 
  $^{5}$ Department of Physics \& Astronomy, Michigan State
  University, East Lansing, MI 48824, USA}

\pagerange{\pageref{firstpage}--\pageref{lastpage}} \pubyear{2012}

\maketitle
\label{firstpage}

\begin{abstract}

  Massive stars provide feedback that shapes the interstellar medium
  of galaxies at all redshifts and their resulting stellar
  populations.  Here we present three adaptive mesh refinement
  radiation hydrodynamics simulations that illustrate the impact of
  momentum transfer from ionising radiation to the absorbing gas on
  star formation in high-redshift dwarf galaxies.  Momentum transfer
  is calculated by solving the radiative transfer equation with a ray
  tracing algorithm that is adaptive in spatial and angular
  coordinates.  We find that momentum input partially affects star
  formation by increasing the turbulent support to a three-dimensional
  rms velocity equal to the circular velocity of early haloes.
  Compared to a calculation that neglects radiation pressure, the star
  formation rate is decreased by a factor of five to $1.8 \times
  10^{-2}$ \hsfr~in a dwarf galaxy with a dark matter and stellar mass
  of $2.0 \times 10^8 \Ms$ and $4.5 \times 10^5 \Ms$, respectively,
  when radiation pressure is included.  Its mean metallicity of
  $10^{-2.1}$ \zsun~is consistent with the observed dwarf galaxy
  luminosity-metallicity relation.  However, what one may naively
  expect from the calculation without radiation pressure, the central
  region of the galaxy overcools and produces a compact, metal-rich
  stellar population with an average metallicity of 0.3 \zsun,
  indicative of an incorrect physical recipe.  In addition to
  photo-heating in \hii~regions, radiation pressure further drives
  dense gas from star forming regions, so supernovae feedback occurs
  in a warmer and more diffuse medium, launching metal-rich outflows.
  Capturing this aspect and a temporal separation between the start of
  radiative and supernova feedback are numerically important in the
  modeling of galaxies to avoid the ``overcooling problem''.  We
  estimate that dust in early low-mass galaxies is unlikely to aid in
  momentum transfer from radiation to the gas.


\end{abstract}


\section{Introduction}

Stellar radiation from massive stars and their supernova (SN)
explosions significantly alter the surrounding interstellar medium
(ISM) and, thus, subsequent star formation in the host galaxy
\citep[for a review, see][]{McKee07}.  On a grander scale, their input
can drive large-scale outflows that increase the entropy and
metallicity of the intergalactic medium \citep[][for a
  review]{Davies98, Aguirre01a, Benson10_Review}.  To date, the
primary feedback mechanism in galaxy formation simulations originate
from SNe, which is implemented by locally injecting thermal and
kinetic energy and metals \citep{Cen92}.  There are various relevant
physics, such as photoheating \citep[e.g.][]{Gnedin00}, momentum input
from radiation \citep[e.g.][]{Haehnelt95}, cosmic rays
\citep[e.g.][]{Jubelgas08}, and magnetic fields \citep[e.g.][]{Wang09,
  Kotarba11}, that can play an important role in star and galaxy
formation.  Numerical simulations are a useful tool in exploring the
impacts of these physical processes in galaxy formation.  But only
recently, feedback mechanisms other than SNe have begun to appear in
cosmological simulations, sometimes in a phenomenological manner,
because of resolution and algorithmic limitations.

Simulating the first low-mass galaxies with the correct physical model
is crucial because they are the building blocks for all subsequent
galaxies, and it is clear that stellar radiation from them were
predominately responsible for reionisation \citep[e.g.][]{Shapiro86,
  Fan06, Bouwens11_Reion}.  The most luminous of these ``reionisers''
have been detected at $6 < z < 10$ in the \textit{Hubble Ultra Deep
  Field} (HUDF) and from ground-based campaigns
\citep[e.g.][]{Bouwens08, McLure10, Finkelstein10, Ouchi10}.  Even
lower mass and higher redshift galaxies will be detected with the
\textit{Atacama Large Millimeter Array} (ALMA) and the \textit{James
  Webb Space Telescope} (JWST), providing additional constraints on
early galaxy formation, reionisation, and the early metal enrichment
of the intergalactic medium.  Closer to home, nearby dwarf galaxies
provide clues to their formation, where all of them formed stars at or
before reionisation \citep{Grebel03}, in their [$\alpha$/Fe] versus
[Fe/H] evolution \citep{Tolstoy09}, metallicity distribution functions
\citep{Kirby11}, metallicity radial gradients \citep{Tolstoy04}, star
formation histories \citep{Monelli10}, and individual abundance
patterns \citep{Frebel10_Spec, Caffau11}.  These data from the
high-redshift and the local universe provide important constraints on
numerical models of star formation and feedback.

The first generation of galaxies with $10^8 \la M_{\rm halo}/\Ms \la
10^{10}$ \citep{Bromm11} are an excellent cosmological testbed to
numerically investigate the impact of each physical process.  Compared
to simulating larger galaxies, they present a relatively clean
scenario with few instances of prior star formation that sets the
``initial conditions'' for galaxy formation.  They are also small
enough so it is computationally feasible to achieve high enough
resolution to well resolve star forming regions.  Their progenitors
may have hosted several metal-free and massive (Pop III) stars,
enriching the IGM and pre-heating the gas to thousands of degrees,
which then re-accretes into the dark matter (DM) halo and forms a
dwarf galaxy \citep{Wise08_Gal, Greif10}.  In this paper, we
investigate the impact of radiative cooling from metals and a
\hh-dissociating radiation background, but the main focus of this work
is the effect of momentum transfer from ionising radiation,
i.e. \textit{radiation pressure}, on star formation in such galaxies.

Regulating star formation by radiation pressure is not a new idea,
especially in the Milky Way and present-day star formation
\citep{Shu91, Ferrara93, Li96, Matzner99}.  However,
\citet{Haehnelt95} was the first to apply this idea to galaxy
formation.  He found that radiation pressure could be the dominant
feedback process within $\sim$110~pc in DM haloes with masses $\la 10^9
- 10^{10} \Ms$.  He suggested that it could be more important than SNe
heating in driving outflows from such early dwarf galaxies.  In turn,
this would lower their gas fractions and alter their predicted
mass-to-light ratios.  In larger galaxies, it was unclear whether
momentum deposition could drive outflows through ionising radiation
alone.  However, UV radiation can be absorbed by dust and be
re-emitted many times, increasing the momentum transfer to the
absorber.  Without making the assumption that gas dynamics are coupled
to dust grains, radiation pressure from massive stars were found to
drive dust outflows in cosmological simulations, enriching the IGM to
the C and Si abundances found in $z=3$ quasar absorption lines
\citep{Aguirre01a, Aguirre01b, Aguirre01c}.  \citet{Murray05} and
\citet{Thompson05} investigated the effects of radiation pressure in
driving galactic outflows from starburst galaxies and regulating star
formation by reducing the gas supply for further star formation and BH
growth.  They found that it could explain the $M_{\rm BH}-\sigma$
relation for early-type galaxies, where $\sigma$ is the stellar
velocity dispersion \citep{Ferrarese00, Gebhardt00, Tremaine02}.  They
also found that momentum-driven winds could be launched if the opacity
$\kappa \ga 6 \; \textrm{cm}^2 \; \textrm{g}^{-1}$ in a starburst galaxy,
assuming that momentum transfer is increased by a factor of $\tau_{\rm
  IR} - 1$, where $\tau_{\rm IR}$ is the optical depth in the infrared
(IR).

These works led to various implementations of momentum transfer in
numerical studies of galaxy formation, usually in the form of
``kicking'' particles with some wind velocity from star-forming
regions \citep[e.g.][]{Springel03b, Oppenheimer08, Sales10,
  Hopkins11_RP}.  The early implementations decoupled these wind
particles from the hydrodynamics to create outflows; otherwise, they
would be confined to the galaxy.  Recently, \citet{Krumholz12_RP}
investigated the parsec-scale effects of momentum transfer in
radiation hydrodynamics simulations and compared them to the
aforementioned subgrid models.  They found that most of the radiation
is not trapped because the optically-thick shell breaks up from a
radiation Rayleigh-Taylor instability, i.e. momentum transfer does not
scale as $\tau_{\rm IR}$.  This may lead to an overestimate of
momentum transfer in the subgrid models when the radiation source
shines above 10 per cent of its Eddington limit.

Here we present results that demonstrate the effects of momentum
transfer from ionising radiation on dwarf galaxy formation, using
cosmological radiation hydrodynamics simulations.  The amount of
momentum deposition is directly calculated from our radiation
transport solver \moray~\citep{Wise11_Moray}.  In the next section, we
outline some analytics that govern the dynamics of outflows and
turbulence created by radiation pressure.  Then in
Sec. \ref{sec:setup}, we describe the numerical simulation setup and
algorithms.  In Sec. \ref{sec:results}, we present the effects of
radiation pressure on star formation and metal enrichment in dwarf
galaxies, preventing the overcooling problem seen often in galaxy
formation simulations.  We discuss possible constraints on
high-redshift dwarf galaxies and then focus on differences in
feedback implementations in simulations in Sec. \ref{sec:discuss}.  We
finish that section with an estimate of momentum transfer onto dust
grains in dwarf galaxies.  We summarise our findings in
Sec. \ref{sec:conclusions}.

\section{Momentum transfer from ionising radiation}
\label{sec:anyl}


First we review some basic analytical arguments on momentum transfer
from ionising radiation.  We take a similar approach as
\citet{Haehnelt95} and \citet{Murray05} and use an Eddington-like
argument on an entire galaxy, equating the radiation pressure from
ionising stellar radiation to the gravitational potential of the host
dark matter halo.  Its application to local star formation and
feedback has been extensively studied \citep[][and references
  therein]{McKee07}.  But can we apply the same principles to an
entire dwarf galaxy?  How will the energetics differ if the momentum
transfer originates from ionising radiation instead of protostellar
winds or Thomson scattering?

In the traditional Eddington approach \citep{Eddington26}, Thomson
scattering is the source of radiation pressure.  The Eddington
luminosity limit is
\begin{equation}
  L_{\rm edd} = \frac{4 \pi G M m_p c}{\sigma_{\rm T}},
\end{equation}
where $M$ is the mass of object, $m_p$ is the proton mass, and
$\sigma_{\rm T} = 6.65 \times 10^{-25} \;\mathrm{cm}^2$ is the Thomson
cross-section.  However, the absorption cross-section for hydrogen
$\sigma_{\rm HI} = 6.35 \times 10^{-18} \;\mathrm{cm}^2$ at 1 Ryd,
which is $\sim 10^7$ times larger than the Thomson cross-section.
However, it should be noted that a spectrum-weighted cross-section is
somewhat lower because $\sigma_{\rm HI} \propto \nu^{-3}$.  Thus
momentum transfer during the absorption of ionising photons might
significantly affect the large-scale dynamics in the galaxy.

For illustrative purposes, we focus on a single expanding \hii~region
with hydrogen only that is embedded in a DM halo.  A shock front forms
at the edge of the \hii~region as when the ionisation front becomes
D-type \citep{Osterbrock89}, i.e., the expansion speed of the
\hii~region slows below the sound speed of the ionised medium.  If the
ionisation front coincides with the shock front, the optical depth
from the radiation source will exceed unity within the shock.  We can
simplify the calculation by assuming that the \hii~region is optically
thin and all of the ionising radiation momentum is transferred to the
optically-thick shock.

\citet{Murray05} considered the spherically symmetric and
self-gravitating case where the stellar system is contained within a
halo with a gas fraction $f_{\rm gas}$.  Their main focus was on
nuclear starbursts and the resulting outflows and explaining the
$M-\sigma$ relationship \citep{Ferrarese00, Gebhardt00, Tremaine02}.
They neglected thermal pressure forces from the \hii~region because it
is negligible for a stellar system that is centered in the
gravitational potential of halos that host $L \ga L^*$ galaxies.  For
a central \textit{ionising} luminosity $L$ where each photon is
absorbed once, the momentum equation is
\begin{equation}
  \label{eqn:p1}
  \frac{dP}{dt} = - \frac{G M(r) M_{\rm gas}(r)}{r^2} + \frac{L}{c}.
\end{equation}
In the case where photons are absorbed and re-emitted many times in a
dusty medium, the second term can become important.
\citeauthor{Murray05} showed that for a luminosity $L \ga 4 f_{\rm
  gas} \sigma^4 c / G$, the gas moves outward driven by pure radiation
pressure, where $\sigma$ is the halo velocity dispersion and all
radiation is assumed to be absorbed.

\begin{figure}
  \plotone{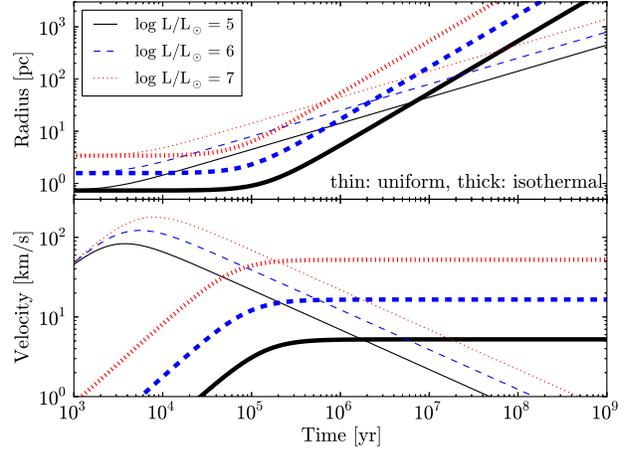}
  \caption{\label{fig:anyl} Radius (top) and velocity (bottom) of an
    expanding optically thick shell driven by momentum transfer from
    ionising radiation for a central luminosity of $L = 10^5, 10^6,
    10^7 \lsun$ with an average ionising photon energy of 20 eV.}
\end{figure}

For a dwarf galaxy with $\sigma = 20 \kms$ and the cosmic baryon
fraction, this critical luminosity is on the order of $10^{42}$ erg
s$^{-1}$ ($10^{8.4} \lsun$), equivalent to a mass-to-light ratio on
the order of unity for a Salpeter IMF.  The gravitational potential in
dwarf galaxies with $V_c \la 30 \kms$ cannot prevent the expansion of
the dense shell surrounding \hii~regions into the IGM.  Furthermore,
star formation is not necessarily concentrated in the centre of these
small irregular galaxies \citep{Tolstoy09}.  If the \hii~region is
pressure-supported \citep[e.g., see][]{Ferrara93}, then we can neglect
the gravitational force term in Eq. \ref{eqn:p1}.  Taking $L$ to be
time-independent and instantaneously non-zero at $t>0$, the momentum
equation for the optically-thick shell is simply $P(t) = Lt/c$.  We
consider the spherically-symmetric case where the shell sweeps all of
the ISM with a density $\rho = \rho(r)$ in its path, and it moves at a
velocity
\begin{equation}
  v(t) = \frac{Lt}{cM_{\rm shell}} = \frac{Lt}{c} \left[ 4\pi
    \int_0^r r^{\prime 2} \rho(r^\prime) dr^\prime \right]^{-1}.
\end{equation}
Solving for radius and velocity with a homogeneous medium $\rho(r) =
\rho_0$ and isothermal density profile $\rho(r) = \rho_0 (r/r_0)^{-2}$, we
obtain
\begin{equation}
  \label{eqn:rshell}
  r(t) = \left\{ \begin{array}{r@{\quad}r}
    \left( r_i^4 + 2A t^2 \right)^{1/4} & (\rho = \rho_0)\\
    \left( r_i^2 + A t^2 / 3 r_0^2 \right)^{1/2} & (\rho
    \propto r^{-2})\\
  \end{array} \right.
\end{equation}
\begin{equation}
  \label{eqn:vshell}
  v(t) = \left\{ \begin{array}{r@{\quad}r}
    t A \left( r_i^4 + 2A t^2 \right)^{-3/4} & (\rho = \rho_0)\\
    t (A / 3 r_0^2) \left( r_i^2 +
    A t^2 / 3 r_0^2 \right)^{-1/2} & (\rho \propto r^{-2})
  \end{array} \right.,
\end{equation}
where $A \equiv 3L /4 \pi \rho_0 c$, and $r_i$ is the radius where the
optically thick shell first forms.  The initial Str\"omgren sphere
radius is calculated by equating the ionising photon luminosity
$\dot{N}_\gamma = L / \eavg$ to the number of recombinations, so that
\begin{equation}
  r_i = \left( \frac{3 \dot{N}_\gamma}{4 \pi n_0^2 \alpha_B} \right)^{1/3},
\end{equation}
given an initial absorber number density $n_0$ of the star-forming
region.  Here \eavg~is the average ionising photon energy, and
$\alpha_B \approx 2.5 \times 10^{-13} \; \textrm{cm}^3 \;
\textrm{s}^{-1}$ is the Case B recombination rate at $T = 10^4$ K.
Molecular complexes are very clumpy and, in the Galactic plane, they
have an average density in the range of $10^2 - 10^4 \; \cubecm$
\citep{Bergin07}.  For a representative stellar cluster,
\begin{equation}
  r_i = 1.57 \; L_6^{1/3} E_{\gamma, 20}^{-1/3} n_3^{-1/3} \;
  \textrm{pc},
\end{equation}
where $L_6 = L/10^6 \lsun$, $E_{\gamma, 20} = E_\gamma/20 \:
\textrm{eV}$, and $n_3 = n_0/10^3 \; \textrm{cm}^{-3}$.  We plot
$r(t)$ and $v(t)$ for $L = 10^{5}, 10^{6}, 10^{7} \lsun$, $\rho_0 = 1
\; \textrm{cm}^{-3}$, and \eavg~= 20 eV in Figure \ref{fig:anyl} for a
uniform and a isothermal density profile.  We set $r_0 = 15 \;
\textrm{pc}$ so that both the uniform and isothermal sphere with
radius $3r_0$ contains $10^4 \Ms$.  In a uniform medium, the shell
rapidly accelerates to its maximum velocity $\sim$ 100~\kms~after
$10^4$ yr.  Its velocity decreases as the shell accumulates material
from the ISM.  It eventually reaches 1 kpc after 1 Gyr for the $10^6
\lsun$ case.  In the isothermal sphere, the shell accelerates as it
runs down the density slope, reaching 1 kpc after 50 Myr.  The shell
reaches a constant velocity $v_s = (L/4 \pi \rho_0 c r_0^2)^{1/2} \sim
13 L_6 \kms$ after about a Myr \citep{Bally83, Shu91, Matzner99} with
a fraction $f = 5.0 \times 10^{-4} L_6^{-0.5}$ of the radiation energy
being transferred to the kinetic energy of the shell.

This ideal model will break down as the shell when gravitational
forces from the DM halo becomes larger than $L/c$ and the thermal
pressure inside the \hii~region.  However if parts of the shell
fragment because of ionisation front or thin-shell instabilities, the
diffuse gas in the openings of the shell will experience a greater
acceleration, i.e. champagne flows \citep[e.g.][]{Yorke83}.  These
radiation pressure driven champagne flows are a possible mechanism for
enhancing mass outflow velocities in dwarf galaxies in this model.

The terminal velocity for the shell traveling down a $r^{-2}$ density
profile is comparable to the circular velocity of the DM halo.
Radiative pressure driven winds in dwarf galaxies may contribute more
to the driving of turbulence instead of outflows when the shells are
contained within the halo.  Collisions between surviving swept-up
shells can drive turbulence in the star-forming region.  This has been
shown to be likely in present-day star formation
\citep[e.g.][]{Norman80, Franco83, Krumholz06, Matzner07}.  This added
turbulent support may help in regulating star formation by preventing
a catastrophic central collapse in a galaxy.  

We can understand the evolution of turbulence by investigating the
vorticity equation,
\begin{equation}
  \label{eqn:vort}
  \frac{D\bomega}{Dt} = -\bomega \nabla \cdot \mathbf{v} -
  \frac{\nabla \mathcal{P} \times \nabla \rho}{\rho^2} + \nu \nabla^2
  \bomega,
\end{equation}
where $D/Dt$ is the fluid derivative,
$\mathcal{P}$ is pressure, and $\nu$ is the viscosity.  In our
analysis, we consider non-viscous fluids and set $\nu = 0$.  The first
term describes the stretching of vorticity; the second term comes from
non-barotropic flows, i.e. $\mathcal{P} \ne \mathcal{P}(\rho)$, and
the last term accounts for the dissipation of turbulence through
viscous forces.  Non-barotropic flows occur at and near shock fronts
and have been found to be the main process that sustains vorticity in
several situations.  For example, virial shocks and internal weak
shocks have been found to drive turbulence during halo virialization
\citep{Wise07, Greif08}.  Furthermore, non-thermal pressure, such as
radiation pressure or cosmic rays, contributes to the generation of
vorticity when its gradient (acceleration) is not aligned with the
density gradient.  To determine the importance of radiation pressure
in outflows and sustaining turbulent motions, we use cosmological
radiation hydrodynamics simulations that focus on dwarf galaxy
formation, which we describe next.

\section{Cosmological Simulations}
\label{sec:setup}


We expand upon the simulation presented in \citet[][hereafter Paper
  I]{Wise12_Galaxy} that focused on the transition from Population III
to II star formation and the effects of radiative and supernova
feedback on early dwarf galaxy formation.  In this work, we compare
two new simulations that add radiative cooling from metals, an
\hh-dissociating radiation background, and momentum transfer from
ionising radiation to the original physics set.  These simulations
were run with the adaptive mesh refinement (AMR) code
\enzo~v2.0\footnote{\texttt{enzo.googlecode.com, changeset
    03a72f4f189a}} \citep{BryanNorman1997, OShea2004}.  It uses an
$N$-body adaptive particle-mesh solver \citep{Efstathiou85,
  Couchman91, BryanNorman1997} to follow the DM dynamics.  It solves
the hydrodynamical equation using the second-order accurate piecewise
parabolic method \citep{Woodward84, Bryan95}, while a Riemann solver
ensures accurate shock capturing with minimal viscosity.  We use the
recently added HLLC Riemann solver \citep{Toro94_HLLC} for additional
stability in strong shocks and rarefaction waves.

We use a simulation box of 1 Mpc on a side with a base resolution of
$256^3$, resulting in a DM mass resolution of 1840 \Ms.  This allows
us to resolve $2 \times 10^5 \Ms$ DM haloes that can cool by \hh~in the
gas phase \citep{Tegmark97} by at least 100 particles.  We refine the
grid on baryon overdensities of $3 \times 2^{-0.2l}$, where $l$ is the
AMR level, resulting in a super-Lagrangian behaviour \citep[also
  see][]{OShea08}.  We also refine on a DM overdensity of three and
always resolve the local Jeans length by at least 4 cells, avoiding
artificial fragmentation during gaseous collapses \citep{Truelove97}.
If any of these criteria are met in a single cell, it is flagged for
further spatial refinement.

We initialise the simulations with \textsc{grafic}
\citep{Bertschinger01} at $z = 130$ and use the cosmological
parameters from the 7-year WMAP $\Lambda$CDM+SZ+LENS best fit
\citep{WMAP7}: $\Omega_M = 0.266$, $\Omega_\Lambda = 0.734$, $\Omega_b
= 0.0449$, $h = 0.71$, $\sigma_8 = 0.81$, and $n = 0.963$ with the
variables having their usual definitions.  We use a maximum refinement
level of $l = 12$, resulting in a maximal comoving resolution of 1 pc.
We do not smooth the DM mass field at any level, which is done to
reduce artificial gas heating from DM particles in simulations that
have regions that are baryon dominated \citep[e.g.][]{ABN02}.  The
adaptive particle-mesh solver has a force resolution of two cell
widths of a given AMR grid.  We compare the simulations at $z=8.1$,
which is when the metal cooling simulation reaches an ionised volume
fraction of 96 percent.

\subsection{Star formation}

We distinguish Pop II and Pop III SF by the total metallicity of the
densest cell in the molecular cloud.  Pop II stars are formed if [Z/H]
$> -4$, and Pop III stars are formed otherwise.  We do not consider
hypothetical Pop III.2 stars or intermediate mass stars from
CMB-limited cooling.  Here we outline how we treat Pop II and III SF
in our simulations; for more details and justifications, we refer the
reader to \citet{Wise12_Galaxy}.

\subsubsection{Population III star formation}

We use the same Pop III SF model as \citet{Wise08_Gal} where each star
particle represents a single star.  In this model, a star particle
forms when a cell has all of the following criteria:
\begin{enumerate}
\item an overdensity of $5 \times 10^5$ ($\sim 10^3\; \cubecm$ at
  $z=10$),
\item a converging gas flow ($\nabla \cdot \mathbf{v}_{\rm gas} < 0$),
  and
\item a molecular hydrogen fraction $f_{\rm H2} > 5 \times 10^{-4}$.
\end{enumerate}
These physical conditions are typical of collapsing metal-free
molecular clouds $\sim 10$ Myr before the birth of a Pop III
main-sequence star \citep{ABN02, OShea07a}.  This is approximately the
end of the initial free-fall phase, the so-called ``loitering phase'',
where the minimum temperature at this point characterises the
fragmentation mass scale \citep[e.g.][]{Omukai03}.  This prescription
is similar to the \citet{Cen92} method but removes the Jeans unstable
($M_{\rm gas} > M_J$) and cooling timescale ($t_{\rm cool} < t_{\rm
  dyn}$) requirements.  We do not consider the former criterion
because it is not applicable to simulations that resolve the Jeans
length at all times.  The molecular hydrogen fraction requirement
effectively constrains star formation to cooling molecular clouds,
where the \hh~formation rate is significantly larger than the
dissociation rate from a Lyman-Werner radiation field.

If multiple cells meet the star particle formation criteria within 1
pc, we form one Pop III star particle with the centre of mass of these
flagged cells to ensure that one massive star is created per
metal-free molecular cloud.  We randomly sample from an IMF with a
functional form of
\begin{equation}
\label{eqn:imf}
f(\log M) dM = M^{-1.3} \exp\left[-\left(\frac{M_{\rm
      char}}{M}\right)^{1.6}\right] dM
\end{equation}
to determine the stellar mass.  Above $M_{\rm char}$, it behaves as a
Salpeter IMF but is exponentially cutoff below that mass
\citep{Chabrier03}.  After the star particle forms and its mass is
determined, an equal amount of gas is removed from the computational
grid in a sphere that contains twice the stellar mass and is centred
on the star particle.  At this stage of the collapse, a sphere
enclosing $200 \Ms$ has a radius of 1--2 pc and has approximately 200
cells.  The star particle acquires the mass-weighted velocity of the
gas contained in this sphere.

\subsubsection{Pop II star formation}

We treat Pop II star formation with the same prescription as
\citet{Wise09}, which is similar to the Pop III prescription but
without the minimum \hh~fraction requirement.  This is removed because
the metal-enriched gas can efficiently cool even in the presence of a
strong UV radiation field \citep[e.g.][]{Safranek10}.  To ensure the
volume is cooling, we restrict star formation to gas with temperatures
$T < 1000$ K.  Unlike Pop III star particles that represent individual
stars, Pop II star particles represent a star cluster of some total
mass and an assumed Salpeter IMF.

Once a cell meets these criteria, the prescription searches outward
with increasing radius for the boundary of the molecular cloud,
centred on the most dense cell.  Here, we define a molecular cloud as
a sphere with a dynamical time $t_{\rm dyn} = 3$ Myr (corresponding to
an average density $\bar{\rho}_{\rm MC} \simeq 500\mu\; \cubecm$) and
a radius $R_{\rm MC}$, where $\mu$ is the mean molecular weight.  This
sphere typically has a radius of 6 pc (increasing as $M_{\rm
  MC}^{1/3}$) for the smallest molecular clouds and 2,500 cells.  Once
the sphere radius is found, a fraction $c_\star = 0.07 f_{\rm cold}$
of the cold gas ($T < 10^3$ K) inside the sphere is converted into a
star particle with mass $m_\star = c_\star (4\pi/3) \bar{\rho}_{\rm
  MC} R_{\rm MC}^3$.  This treatment of cold gas accretion is similar
to a star formation model with a multi-phase subgrid model
\citep{Springel03b} but is employed in a simulation that can resolve
the multi-phase interstellar stellar medium (ISM).  After the star
particle is created, we replace the sphere with a uniform density
$\rho_{\rm MC} = (1 - c_\star) / (Gt_{\rm dyn}^2)$ and temperature $T
= 10^4$ K, which approximates the initial stages of an \hii~region.

\subsection{Stellar feedback}

The radiation field is evolved with adaptive ray tracing
\citep{Abel02_RT, Wise11_Moray} that is based on the HEALPix framework
\citep{HEALPix} and is coupled self-consistently to the hydrodynamics.
Each star particle is a point source of hydrogen ionising radiation
with the ionising luminosity equally split between 48 initial rays
(HEALPix level 2).  We use a mono-chromatic spectrum for the radiation
with the energy $E_{\rm ph}$, the spectral shape weighted photon
energy.  For a cosmological simulation that focuses on galaxies, this
simplification does not significantly affect the overall galactic
dynamics \citep[see Sec. 6.3 in][]{Wise11_Moray}.  We do not consider
\hei~and \heii~ionising radiation.  As the rays propagate from the
source or into a high resolution AMR grid, they are adaptively split
into 4 child rays, increasing the angular resolution of the solution,
when the solid angle of the ray $\theta = 4\pi/(12 \times 4^{L})$ is
larger than 1/3 of the cell area, where $L$ is the HEALPix level.  We
limit the ray splitting to a maximum HEALPix level of 13, or
equivalently $8.05 \times 10^8$ rays per source.  We model the
\hh~dissociating radiation with an optically-thin, inverse square
profile, centred on all Pop II and III star particles.

\subsubsection{Population III stellar feedback}

We use mass-dependent hydrogen ionising and LW photon luminosities and
lifetimes of the Pop III stars from \citet{Schaerer02}.  We use a
mass-independent photon energy $E_{\rm ph} = 29.6$ eV, appropriate for
the near-constant $10^5$ K surface temperatures of Pop III stars.
They die as Type II SNe if $11 \le M_\star/\Ms \le 40$
\Ms~\citep{Woosley95, Nomoto06} and as PISNe if they are in the mass
range 140--260 \Ms~\citep{Heger03}, where $M_\star$ is the stellar
mass.  For normal Type II SNe between 11--20 \Ms, we use an explosion
energy of \tento{51}~erg and a linear fit to the metal ejecta mass
calculated in \citet{Nomoto06},
\begin{equation}
  \label{eqn:typeii}
  M_{\rm Z}/\Ms = 0.1077 + 0.3383 \times (M_\star/\Ms - 11)
\end{equation}
or equivalently, the metal yield fraction
\begin{equation}
  y = 0.3383 - 3.614 M_\star^{-1}.
\end{equation}
We model the SNe of stars with $20 \le M_\star/\Ms \le 40$ as
hypernova with the energies and ejecta masses ($y \sim 0.15 - 0.2$)
also taken from \citeauthor{Nomoto06}, linearly interpolating their
results to $M_\star$.  For PISNe, we use the explosion energy from
\citet{Heger02}, where we fit the following function to their models,
\begin{equation}
  \label{eqn:pisn}
  E_{\rm PISN} = 10^{51} \times [5.0 + 1.304 (M_{\rm He}/\Ms - 64)] \;
  \mathrm{erg},
\end{equation}
where $M_{\rm He} = (13/24) \times (M_\star - 20) \Ms$ is the helium
core (and equivalently the metal ejecta) mass and $M_\star$ is the
stellar mass.  If the stellar mass is outside of these ranges, then an
inert, collisionless black hole (BH) particle is created.

The blastwave is modelled by injecting the explosion energy and ejecta
mass into a sphere of 10 pc, smoothed at its surface to improve
numerical stability \citep{Wise08_Gal}.  Because we resolve the
blastwave relatively well with several cells across at its
initialization, the thermal energy is converted into kinetic energy
and agrees with the Sedov-Taylor solution \citep[e.g.][]{Greif07}.

\subsubsection{Population II stellar feedback}

Pop II star particles emit 6000 hydrogen ionising photons per stellar
baryon averaged over their lifetime and $E_{\rm ph} = 21.6$ eV,
appropriate for a [Z/H] = $-1.3$ population \citep{Schaerer03}.  The
star particles live for 20 Myr, the maximum lifetime of an OB star.
These stars generate the majority of the ionising radiation and SNe
feedback in stellar clusters, thus we ignore any feedback from lower
mass stars.  By considering a constant luminosity, we may be
underestimating the impact of radiative feedback because, given an
IMF, the total luminosity of a cluster is maximal at early times when
thermal and radiation pressure forces are the greatest, i.e. the edge
of \hii~region is still near the massive stars.

For SN feedback, these stars generate $6.8 \times 10^{48}$ erg
s$^{-1}$ $\Ms^{-1}$ after living for 4 Myr, which is injected into
spheres with a radius of 10 pc.  However, if the resolution of the
grid is less than 10/3 pc, we distribute the energy into a $3^3$ cubic
region surrounding the star particle.  Here, star particles also
return ejected material with mass
\begin{equation}
  \label{eqn:SNmass}
  \Delta m_{\rm ej} = \frac{0.25 \; \Delta t \; M_\star} {t_0 -
    4\;\mathrm{Myr}}, \quad (4 \: \mathrm{Myr} < t - t_{\rm birth} < t_0)
\end{equation}
back to the grid at every timestep on the finest AMR level.  $t_0$ =
20 Myr denotes the lifetime of the star particle.  This ejected gas
has solar metallicity $Z = 0.02$, resulting in a total metal yield $y
= 0.005$.  However, we note that calculations which are calibrated to
Milky Way observations \citep{Madau96} suggest $y \sim 0.02$, and thus
we may be underestimating the Pop II chemical feedback.

\subsection{Simulation variations}

All three simulations start from identical 1 Mpc $256^3$ initial
conditions, have the same refinement criteria, and use the same star
formation and feedback models.  Starting with our reference model in
Paper I, each following simulation builds upon the physics of the
previously described one.

\subsubsection{Reference model}

We use the nine-species (\hi, \hii, \hei, \heii, \heiii, e$^-$, \hh,
\hh$^+$, H$^-$) non-equilibrium chemistry model in
\enzo~\citep{Abel97, Anninos97} and the \hh~cooling rates from
\citet{Glover08_Rates}.  We spatially distinguish metal enrichment
from Pop II and Pop III stars, however they do not contribute to the
radiative cooling rates and act as passive tracer fields for the SNe
ejecta.  We will refer to this simulation as the ``Base'' simulation.
It required 150,000 CPU hours on 512 compute cores to run to $z=8.1$.

\subsubsection{Metal cooling}

In this simulation, we only add the effects of radiative cooling from
fine-structure transitions in metals.  We will refer to this
simulation as the ``MC'' simulation.  We use the method of
\citet{Smith08_Cooling} that is incorporated into \enzo~v2.0.  We
calculate the cooling rates with \textsc{Cloudy 07.02.01}, which has a
sophisticated chemical network that includes all atomic species and
many molecular species\footnote{Because the cooling rates from
  rotational and vibrational \hh~transitions are already included in
  \enzo, thus we do not include \hh~cooling in the \textsc{Cloudy}
  rate table.}, on a logarithmically spaced 4-dimensional grid of (i)
density -- 10$^{-6}$ to 10$^6$ \cubecm, $\Delta$ = 0.25 dex, (ii)
temperature -- 10 to 10$^8$ K, $\Delta = 0.1$ dex, (iii) electron
fraction -- 10$^{-6}$ to 1, $\Delta = 0.25$ dex, and (iv) metallicity
-- 10$^{-6}$ to 1 $\zsun$, $\Delta = 1$ dex, where $\Delta$ is the
spacing between grid points.  We use the CMB radiation spectrum as the
input spectrum.  Dust cooling becomes important at $\ga 10^9 \cubecm$,
which our simulations are not designed to probe, and thus we do not
consider cooling from dust grains.  It required 200,000 CPU hours on
512 compute cores to complete.  Compared to the Base simulation, the
additional compute time comes from ray tracing through a larger
ionised volume, i.e. the rays are not terminated when fully absorbed.

\subsubsection{Radiation pressure and soft UV background}

\begin{figure}
  \plotone{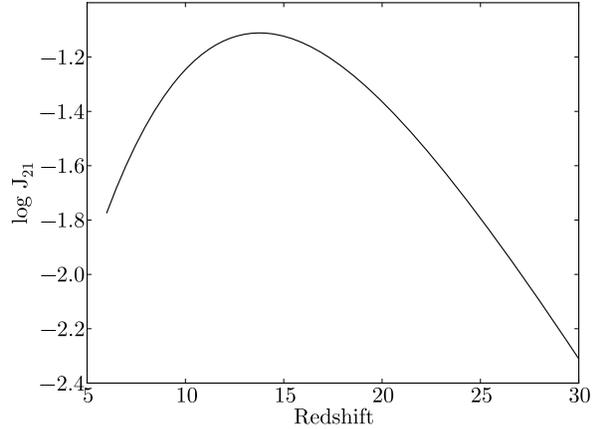}
  \caption{\label{fig:uvb} Time evolution of the Lyman-Werner
    background in units of $J_{21} = 10^{-21}$ \emis, calculated with
    the \citet{Wise05} model.  See eq. (\ref{eqn:uvb}) for its
    functional fit.}
\end{figure}

This simulation adds momentum transfer from ionising radiation to gas
and an \hh-dissociating radiation background.  We will refer to this
simulation as the ``RP'' simulation, and the runtime is similar to the
Base simulation at 150,000 CPU hours.  As the radiation field is
calculated with \moray, the momentum of absorbed ionising radiation
\begin{equation}
d \mathbf{p}_{\rm rp} = \frac{\mathrm{d}P \; E_{\rm ph}}{c}
\hat{\mathbf{r}}
\end{equation}
is transferred to the absorbing medium, where d$P$ is the number of
photons absorbed, $E_{\rm ph}$ is the photon energy, and
$\hat{\mathbf{r}}$ is the normal direction of the radiation.  This
additional momentum further accelerates the gas by
\begin{equation}
  \label{eqn:arp}
  d \mathbf{a}_{\rm ap} = \frac{\mathrm{d}\mathbf{p}_{\rm rp}}{\mathrm{d}t \;
    \rho \; V_{\rm cell}}
\end{equation}
per unit mass, where $\rho$ is the gas density, d$t$ is the radiative
transfer timestep, and $V_{\rm cell}$ is the cell volume.  Currently
we do not consider momentum transfer by radiation trapping when UV
light is absorbed by dust grains and re-emitted in the IR and absorbed
many times.  In principle, this would increase the d$\mathbf{a}_{\rm
  ap}$ by a factor of $f_{\rm trap}$, which we discuss in Section
\ref{sec:discuss_trap}.  The acceleration d$\mathbf{a}_{\rm ap}$ is
then added to the acceleration field in the calculation, for example,
from gravity, in an operator split fashion.

A time-dependent Lyman-Werner (LW) optically thin radiation background
is utilised in this simulation on which the LW radiation from point
sources are added.  LW flux from point sources dominates over the
background within 3 kpc, dissociating \hh~and thus delaying Pop III
star formation in nearby haloes \citep{Machacek01, Wise07_UVB,
  OShea08}.  To calculate its intensity as a function of time, we use
the semi-analytical model of \citet{Wise05}, updated with the 7-year
WMAP parameters and optical depth to Thomson scattering.  The LW
radiation intensity is plotted in Fig. \ref{fig:uvb}.  In this
semi-analytical model, we use a Pop III stellar mass of 100 \Ms, star
formation efficiency of 0.005, and escape fraction of 0.2
\citep{Wise09}.  The intensity decreases after $z \sim 14$ because Pop
II star formation becomes dominant, which produce less LW specific
luminosity than Pop III stars that have surface temperatures of $T =
10^5$ K.  For computational convenience, we fit the background
evolution with the function
  \begin{equation}
    \label{eqn:uvb}
    \log_{10} J_{21}(z) = A + Bz + Cz^2 + Dz^3 + Ez^4,
  \end{equation}
where $(A,B,C,D,E) = (-2.356, 0.4562, -0.02680, 5.882 \times 10^{-4},
-5.056 \times 10^{-6})$ and $J_{21}$ is the specific intensity in
units of $10^{-21}$ \emis.  This fits the model data within 1 per cent
in $6 \le z \le 30$ and is consistent with $J_{21}$ values in
\citet{Trenti09_SFR}.

\section{Results}
\label{sec:results}

We concentrate on the most massive halo in the simulation to
demonstrate the effects of momentum transfer from ionising radiation.
For reasons discussed later, the halo in the MC simulation overcools
and forms stars at high specific SFR, which ionises the entire
simulation volume at $z = 8.1$.  Solving the radiative transfer
equation with \moray~in this simulation becomes computationally
unfeasible in the optically thin limit, so we stop the MC simulation
here and compare the halo in the three simulations that this point.

At redshift 8.1, the halo has a DM mass of $2.0 \times 10^8 \Ms$,
corresponding to a virial temperature $1.7 \times 10^4$ K and circular
velocity of 21 \kms.  This halo experienced a major merger at $z \sim
10$.  It is undergoing a major merger at $z=8.1$ that increases the
virial temperature to $>10^4$ K at which point it can efficiently form
stars.  The gas fraction of the halo $f_g = M_{\rm gas}/M_{\rm tot}$
has recovered to 0.13, previously being depleted to under 5 per cent
in the Pop III hosting progenitors that had its gas expelled by
stellar and supernova feedback.  Here $M_{\rm tot}$ is the total mass
(including DM) of the halo.  At this time, the Base, MC, and RP
simulations have a stellar mass of $2.5 \times 10^5$, $6.7 \times
10^5$, and $4.5 \times 10^5$ \Ms, respectively.  By redshift 7, its
total mass is $1.0 \times 10^9 \Ms$, containing $2.1 \times 10^6 \Ms$
of stars in the Base simulation.  We refer the reader to Paper I for
more details at $z=7$.

\subsection{Comparison of star formation histories}

\begin{figure}
  \plotone{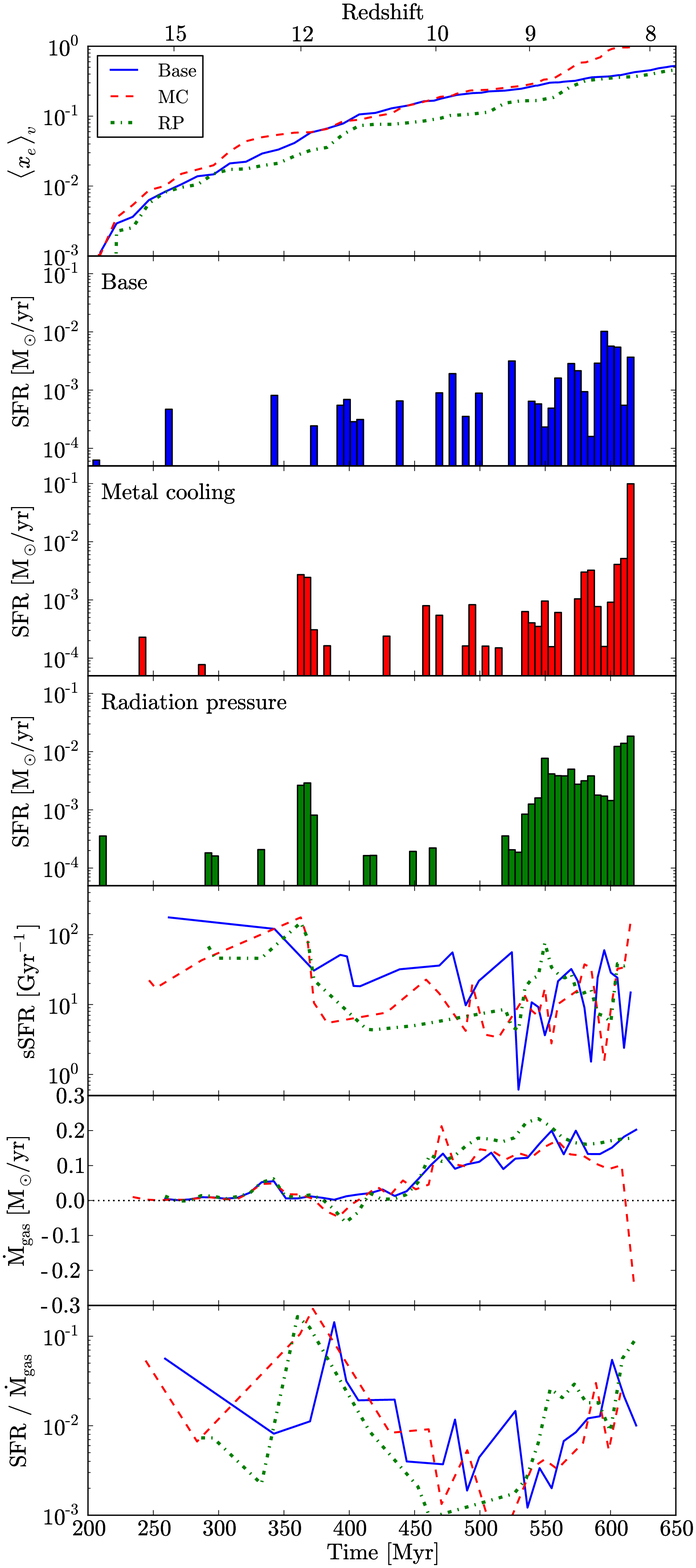}
  \caption{\label{fig:history} The top panel shows the ionised volume
    fraction, while the next three panels show the star formation
    history for the reference, metal cooling, and radiation pressure
    simulations.  The fifth panel displays the specific star formation
    rate for all three simulations.  The sixth panel shows the gas
    accretion rate in the halo, and the bottom panel shows the ratio
    of star formation and gas accretion, which is effectively the star
    formation efficiency.  Including metal cooling results in
    overcooling in the halo interior, efficiently producing stars at a
    rate of $\sim 0.1$ \hsfr.  This single burst ionises the entire
    simulation box.  Radiation pressure from ionising radiation
    regulates star formation to $\sim 0.01$ \hsfr.}
\end{figure}

\begin{figure*}
  \epsscale{2.0}
  \plotone{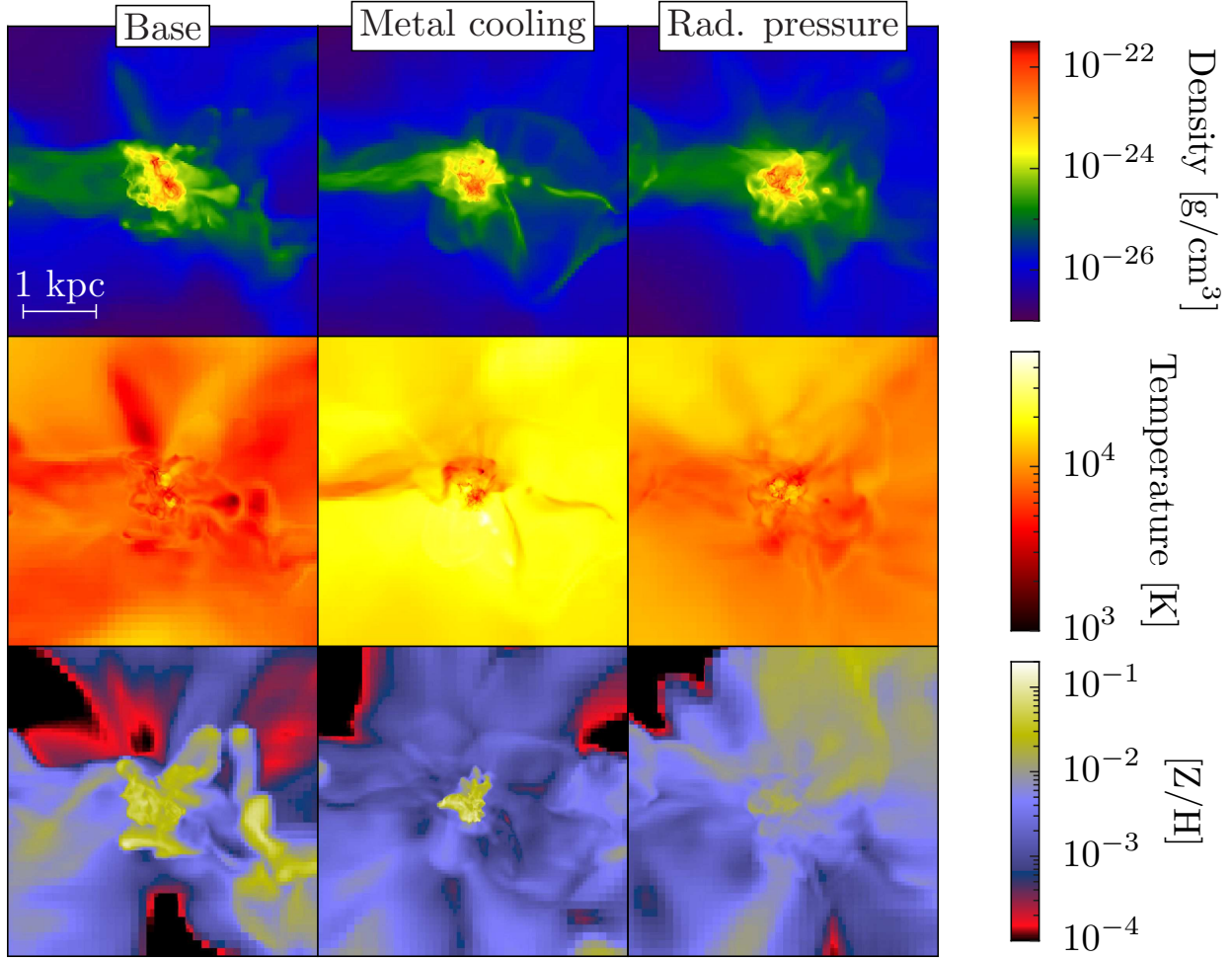}
  \epsscale{1.0}
  \caption{\label{fig:proj} A visual comparison of the most massive
    ($2.3 \times 10^8 \Ms$) halo at $z = 8.1$ in simulations with the
    base physics model (left column), including radiative cooling from
    metal species (middle column), and including radiation pressure
    and an \hh-dissociating background (right column).  The different
    rows contain density-weighted projections of gas density (top),
    temperature (middle), and total metallicity (bottom).  The field
    of view is two virial radii (3.4 kpc).  Comparing the Base and
    Metal Cooling simulations, the metal-rich ejecta cannot expand
    before all of its thermal energy is radiated away, primarily
    through the additional metal cooling.  The additional impetus from
    radiation pressure alleviates this artifact and can regulate star
    formation by driving gas outflows and further entrainment with the
    surrounding medium.}
\end{figure*}

\begin{figure*}
  \epsscale{2.0}
  \plotone{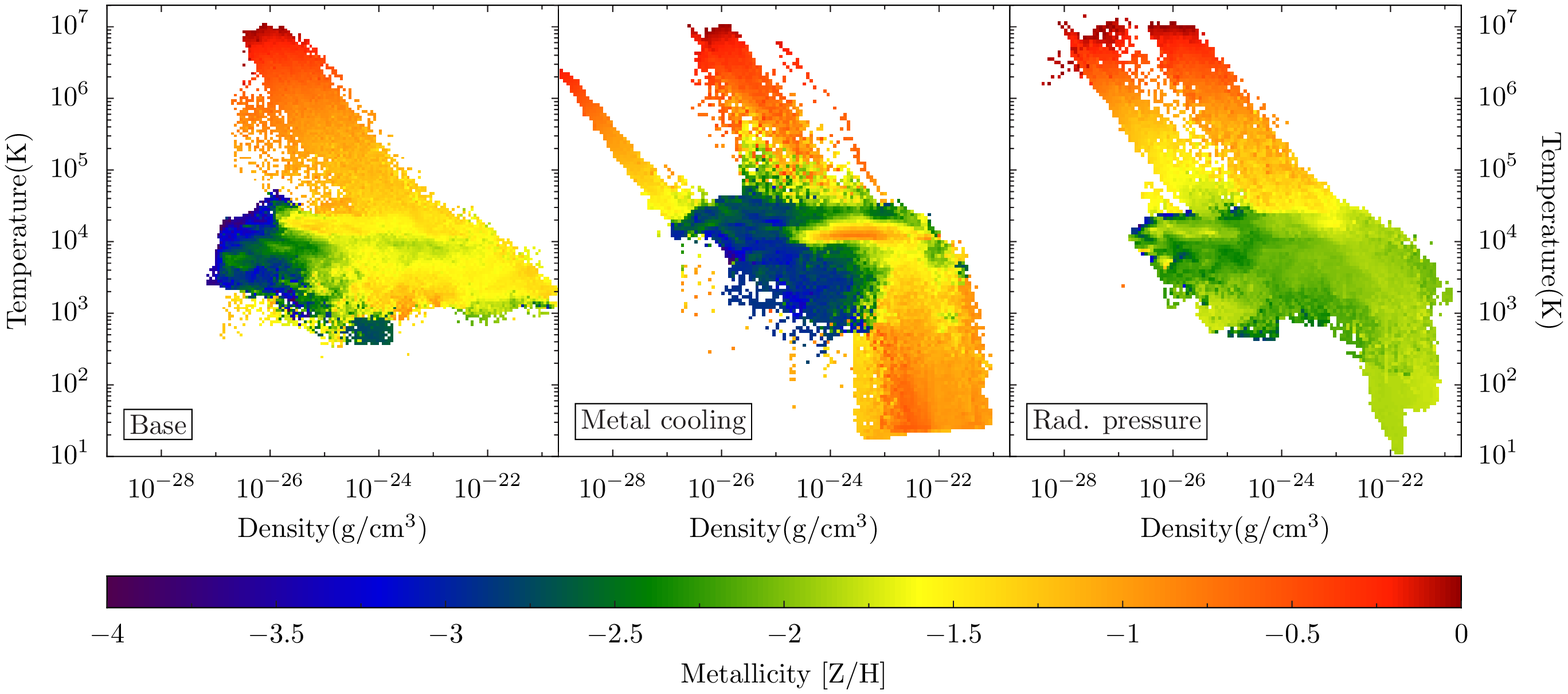}
  \epsscale{1.0}
  \caption{\label{fig:phase}
    Mass-weighted average metallicities as a function of density and
    temperature in the most massive halo at $z = 8.1$ for the Base
    (left), metal cooling (middle), and radiation pressure (right)
    simulations.  When metal cooling is considered, metal-rich gas
    above $3 \times 10^{-23}$ g \cubecm~can cool to the CMB
    temperature.  The majority of gas with $> 10^{-2} \zsun$ is
    restricted to SN remnants (hot ISM phase up to $T = 10^7$ K) and
    cold and dense gas.  The metals are not mixed into the lower
    density ISM.  Radiation pressure (right panel) provides an
    additional impetus to mix the metal-rich ejecta with the entire
    ISM, resulting in nearly all of phase-space being enriched to an
    average of $\sim 10^{-2} \zsun$.}
\end{figure*}

The top panel in Figure \ref{fig:history} shows the evolution of the
ionised volume fraction of all three simulations.  The second, third,
and fourth panels show the star formation rate (SFR).  The specific
SFR (sSFR), the gas accretion rate $\dot{M}_{\rm gas}$, and the ratio
of SFR to $\dot{M}_{\rm gas}$ are shown in the bottom three panels.
At redshift 12, both MC and RP simulations produce metal-enriched
stars after the halo reaches the filtering mass \citep{Hui98,
  Gnedin00, Wise08_Gal} of $\sim 10^7 \Ms$.  The burst in the Base
simulation is slightly delayed and weaker by an order of magnitude
because of the lack of metal cooling.  In all three simulations, about
15 per cent of the accreted gas is converted into stars.  Radiative
and supernova feedback in the stronger bursts of the MC and RP runs
drive a net outflow from the halo, seen in negative values of
$\dot{M}_{\rm gas}$.

The halo continues to accrete from numerous minor mergers of $\sim
10^6 \msun$ haloes until $z = 8.1$.  One notable difference in the RP
simulation is the enhanced gas accretion between redshift 9 and 10.
Here, the UV background suppresses Pop III star formation in
minihaloes by increasing the minimum mass to cool and condense by
\hh~alone \citep{Machacek01, Wise07_UVB, OShea08}.  This results in
more gas-rich mergers, thus increasing the gas accretion rate.  The
halo continues this constant but slow mass buildup until a major
merger at $z = 8.1$.  Before the major merger, the SFRs are similar
between $10^{-2}$ and $10^{-3}$ \hsfr~in the simulations with the RP
simulation sustaining a more consistent SFR mainly because of the
additional gas accretion whereas the other simulations have more a
bursty behavior.  The galaxies convert between 0.1 and 1 per cent of
the accreted gas into stars during this quiescent period before it
reaches a virial temperature of 10$^4$ K.

The most significant difference between the simulations is the star
formation when the halo reaches a virial temperature of $10^4$~K
through a major merger.  The Base simulation has a SFR = $3.7 \times
10^{-3}$ \hsfr~and a sSFR = 15 \ssfr.  In the MC
simulation, the additional radiative cooling provided by the metal
lines prompts a SFR = $9.9 \times 10^{-2}$ \hsfr~and a sSFR = 150
\ssfr, an order of magnitude higher than the Base simulation.  The
radiative and supernovae feedback from these stars drives a net gas
outflow of --0.2 \hsfr~from the halo, as seen in the fourth panel of
Figure \ref{fig:history}, but cannot abate the star formation alone.
Momentum input from massive stars in the RP simulation reduces the SFR
by a factor of five to $1.8 \times 10^{-2}$ \hsfr~and the sSFR to 37
\ssfr.  On average, approximately 3 per cent of the accreted gas forms
stars in the latest star formation period, and the variations are
caused by stellar feedback.  At $z=8.1$, the enhanced star formation
in the MC simulation ionises 96 percent of the simulation volume,
which is in stark comparison with the Base and RP simulations with
ionised fractions of 42 and 36 percent, respectively.  The RP ionised
fraction is slightly lower because of the suppression of Pop III SF
from the LW background.

Observations of star-forming galaxies show a tight relation between
sSFR and stellar mass, and we can compare our results to determine our
most physical model.  Massive ($M_* \ge 10^{10} \Ms$) galaxies at $z
\ge 2$ have $\textrm{sSFR} \propto M_*^{-0.4}$, flattening to an upper
threshold of 2.5 \ssfr~in low-mass galaxies \citep{Karim11}.  At $z =
4-6$, \citet{Stark09} found a strong sSFR-$M_\star$ correlation with a
similarly large scatter \citep[see Figure 1 in][]{Khochfar11}.  In
that work, the average SFR is $\sim 3$ \hsfr~in systems between
$M_\star = 10^8$ and $10^{11}$ \Ms, or equivalently, an average of 30
\ssfr~in stellar systems of $10^8 \Ms$.  Galaxies at $z \ga 7$ also
show a large scatter in sSFR, ranging from 1 to 100 \ssfr~for galaxies
with stellar masses between $10^7$ and $10^9$ \Ms, trending toward
higher sSFRs in smaller systems \citep{Schaerer10, McLure11,
  Khochfar11}.  Do high-redshift galaxies have more gas available for
star formation to sustain sSFR $>$ 2.5 \ssfr, as suggested by the $z
\ga 7$ data?  Does this relation continue to the very smallest
galaxies at high-redshift?  If we assume so, the MC simulation
produces the best match with the other models falling below the
relation.  This goes against intuition in that a more realistic model,
in our case, adding radiation pressure and a LW background, produces
more realistic galaxies.  To further understand this behavior, we must
investigate the gaseous properties and stellar population of the
galaxy.

\subsection{Overcooling and artificially enhanced star formation}

Here we concentrate on the causes behind the enhanced star formation
in the metal cooling simulation when compared to the Base simulation.
Figure \ref{fig:proj} shows density-weighted projections of gas
density, temperature, and total metallicity for the three simulations.
The gas distributions are irregular and have no organised disc
rotation in any of the runs.  There are three DM filaments that
accrete onto the halo, which is typical of rare peaks
\citep[e.g.,][]{Dekel09, Danovich11}.  Only one filament has retained
its gas, which is located to the left of the halo in Figure
\ref{fig:proj}.  The other filaments have been photo-evaporated and
Jeans-smoothed by stellar radiation.  As previously mentioned, the MC
simulation is nearly fully ionised with an average intergalactic
medium (IGM) temperature of $\sim 3 \times 10^4$ K, clearly seen in
the temperature projections.  The additional pressure provided by the
heated IGM compresses the filament in comparison with the Base and RP
simulations.

The spatial metallicity distributions are the most relevant in
understanding the SFR deviations between the simulations.  In the Base
simulation, the metal-rich outflows with [Z/H] between --1 and --2 can
be seen expanding out to a radius of $\sim 1$ kpc.  The MC simulation
produces a factor of three more stars, but the metal-rich ejecta are
restricted in a small volume with a radius of $\sim 0.5$ kpc with a
mean metallicity of [Z/H] = --0.5.  These metal-rich regions are
surrounded by a metal-poor interstellar medium (ISM) with
$-4 < \textrm{[Z/H]} < -2$.

Figure \ref{fig:phase} shows the metallicity as a function of density
and temperature, and it further illustrates the differences that metal
cooling induces.  Without metal cooling in the Base simulation, the
gas can only cool to $\sim 1000$ K through \hh~cooling in the gas
phase.  The metals are mixed to an average of $3 \times 10^{-2} \zsun$
in gas with $\rho > 3 \times 10^{-26} \cubecm$.  The nearly solar
metallicity isobars reaching $10^7$ K are Type II SNe remnants.  When
metal cooling is considered, the relatively metal-rich ($10^{-1}
\zsun$) material is mainly confined to the cold phase of the ISM above
1~\cubecm~and below 1000~K down to the CMB temperature.  This cold,
metal-rich gas mixes little with the surrounding diffuse ISM that has
a mean metallicity of $10^{-3} \zsun$.

Because stars form in cold and dense gas, the stellar metallicities
reflect the gas metallicities in the cold phase.  They can be a
plausibility check as they can be indirectly compared to observed
metallicities in local dwarf galaxies, which have a clear
luminosity-metallicity relationship that increases from [Fe/H] = --2.5
to --1.5 over a luminosity range log$(L/L_\odot)$ = 3--7
\citep{Kirby11}.

\begin{figure}
  \plotone{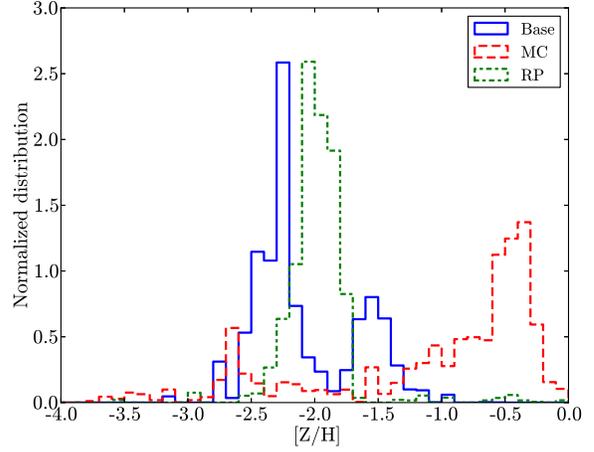}
  \caption{\label{fig:mdf} Normalised stellar metallicity distribution
    functions of the most massive halo for the Base (solid lines),
    metal cooling (dashed), and radiation pressure (dash-dotted)
    simulations at $z=8.1$.  Overcooling in the metal cooling
    simulation creates a stellar population that is nearly solar
    metallicity, whereas the additional mixing with radiation pressure
    shifts the average back to $\sim 10^{-2} \zsun$, agreeing with the
    mass-metallicity relation of galaxies.}
\end{figure}

Figure \ref{fig:mdf} compares the metallicity distribution functions
(MDFs) in the most massive galaxy at $z=8.1$ in each simulation.  The
Base simulation has a bimodal distribution within the range $-2.5 <$
[Z/H] $< -1.0$.  The lower metallicity stars were formed before the
halo reaches $\tvir = 10^4$ K.  Afterward, the stars quickly enrich
the halo to [Z/H] $\sim$ --1.5, which forms stars at similar
metallicities, creating a bimodal MDF.  This process is described in
more detail in Paper I.  A similar enhancement in star formation
occurs in the MC simulation at $\tvir = 10^4$ K but to a greater
extent because of higher cooling rates and thus star formation rates.
Stars primarily form in the inner 500 pc, where the gas is enriched to
[Z/H] $\sim$ --0.5 because the metal-rich ejecta is confined within
these inner regions.

These metallicities are a factor of 30 above the
luminosity-metallicity relationship and suggest that our MC model
produces galaxies that are too luminous and metal-rich.  The extra
radiative cooling from metal species allows for the super-solar SNe
ejecta to lose most of its thermal energy before it launches a
high-velocity outflow, and most of the ejecta are trapped within
the galaxy.  Then this metal-rich material cools again and condenses
into star forming clouds.  This process repeats several times,
creating a runaway effect in overproducing stars, known as the
well-known and studied numerical ``overcooling problem'' that produces
galaxies that are too centrally concentrated \citep[first reported
  by][]{Katz96}.  This was solved by introducing SNe feedback and
conducting the simulations at higher resolutions; however, our
calculations show the same problem at parsec-scale spatial resolution
and despite resolved SNe feedback.

\subsection{Effects of radiation pressure}

We find that momentum transfer from ionising radiation to the
absorbing gas alleviates the overcooling problem seen in the MC
simulation.  Recall that at $z=8.1$, the SFR is reduced by a factor of
five compared to the MC simulation.  To understand this difference, we
focus on its effect on internal gas dynamics and the chemo-thermal
state of the gas, the increased ejecta dispersion from their origin,
and the ensuing star formation.

\subsubsection{Gas dynamics}

\begin{figure}
  \plotone{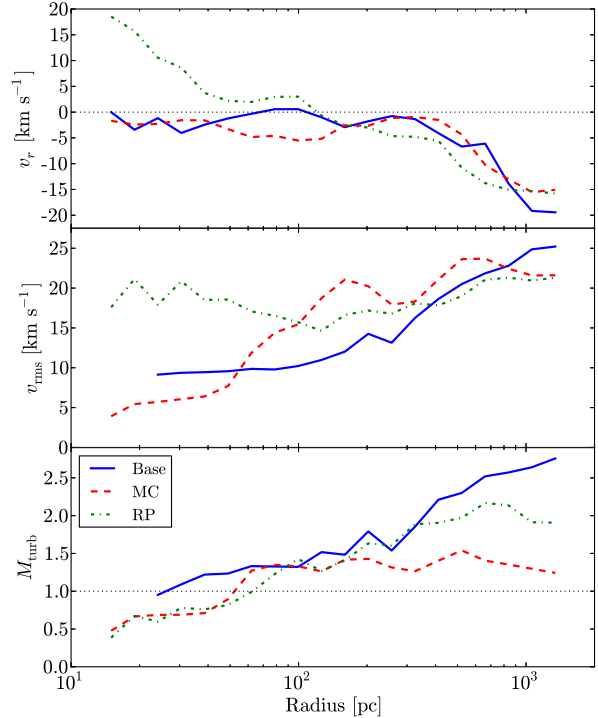}
  \caption{\label{fig:vr} \textit{Top panel:} Mass-weighted radial
    profiles of gas radial velocity of the most massive halo, centred
    on the halo's centre of mass, from the reference (solid), metal
    cooling (dashed), and radiation pressure (dash-dotted)
    simulations.  The Base and MC runs exhibit a zero to small infall
    velocity within 300 pc, whereas adding radiation pressure results
    in an average gas outflow in the inner 100 pc.  \textit{Middle
      panel:} Three-dimensional rms velocity as a function of radius,
    illustrating the turbulent medium created by momentum transfer
    from ionising radiation.  Turbulence motions support the object
    from further collapsing and assist to mix the metal-rich ejecta
    with the metal-poor ISM.  \textit{Bottom panel:} Radial profile of
    the turbulent Mach number $M_{\rm turb}$.  The MC and RP
    simulations exhibit nearly identical $M_{\rm turb}$ values in the
    inner 100 pc, showing that the turbulence decays in the MC run as
    the gas rapidly cools.}
\end{figure}

Unlike the thermal energy produced by SN feedback, momentum must be
conserved and is not lost to radiative processes.  The momentum
transferred to the absorbing gas generates a medium in the central 100
pc that is both turbulent and expanding radially, where the young
stars and their \hii~regions are clustered.  As the D-type shocks
expand with the ionisation fronts, momenta from the absorbed photons
are transferred to the optically thick shock.  Their cumulative action
produces an average outflow in the inner 100 pc, which is visible in
the radial profile of the mass-averaged radial velocity, shown in the
top panel of Figure \ref{fig:vr}.  The radial velocity is calculated
after subtracting the mass-averaged bulk velocity of the halo.  In the
central 15 pc, the mass-averaged radial velocity is 20~\kms, and it
gradually decreases to 2~\kms~at 40 pc.  This weak expansion of the
inner dense gas extends out to 100 pc.  In contrast, the Base and MC
simulations both exhibit weak inflows up to 5~\kms, where SN feedback
is not enough to support the dense central gas.

Within the inner 100 pc, there are 49 young ($< 20$ Myr) star
particles with a total mass of $4.6 \times 10^4 \Ms$.  The interaction
between the expanding shocks associated with their \hii~regions and SN
blastwaves creates and sustains this turbulent medium.  To measure the
turbulent motions, we calculate the three-dimensional rms velocity
\vrms~as a function of radius.  It is computed with respect to the
mass-averaged velocity of each thick shell.  On average, it is $\sim
20 \kms$ and varies little with radius, indicating that turbulence is
widespread throughout the halo.  The turbulence is mildly supersonic
at $r > 100$ pc with turbulent Mach numbers $M_{\rm turb} = \langle
v_{\rm rms} \rangle / \langle c_s \rangle$ decreasing with radius from
two at the virial radius, where $c_s$ is the sound speed and the
angled brackets denote a mass-weighted average in the shell.  This is
nearly the circular velocity of the halo, $V_c = 21 \kms$.  In the
inner 50 pc, $\vrms \sim 20 \kms$, and it slowly decreases to
15~\kms~at 130 pc and increases to 20~\kms~at the virial radius.
Outside 300 pc, radiation pressure has little effect on turbulent
motions, where it is primarily driven by virialization \citep{Wise07,
  Greif08}.  The Base simulation drops to 10~\kms~within $r=100$ pc as
the gas cools and condenses.  The loss of gaseous kinetic energy is
even more apparent with metal cooling, lowering \vrms~to 5~\kms~in the
inner 50 pc.  This is consistent with an overcooling core that is not
pressure supported by either thermal or turbulent energy.

\begin{figure}
  \plotone{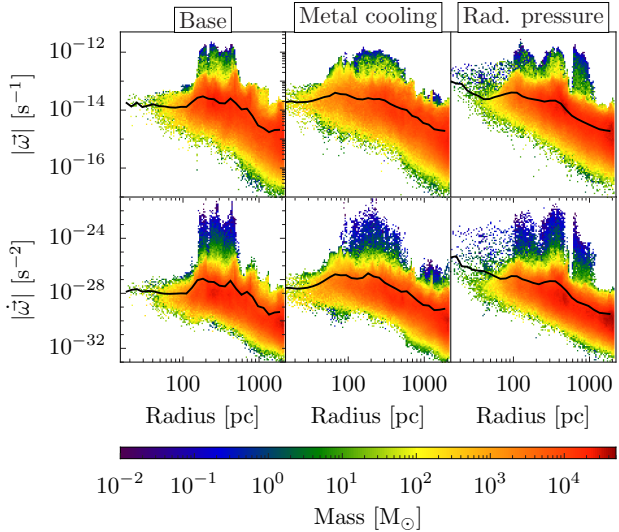}
  \caption{\label{fig:vort} \textit{Top row:} Mass-weighted histograms
    of the magnitude of vorticity for the Base, MC, and RP runs from
    left to right.  The mass-weighted radial profile of vorticity is
    shown by the solid line.  \textit{Bottom row:} Total time
    derivative of vorticity, i.e. right hand side of Equation
    (\ref{eqn:vort}).  Within the inner $\sim$70 pc, vorticity is
    generated at a higher rate in the RP simulations than the Base and
    MC cases, leading to a more turbulent medium that is also depicted
    in the rms velocity (Figure \ref{fig:vr}).}
\end{figure}

Interestingly, the turbulent Mach numbers, shown in the bottom panel
of Figure \ref{fig:vr}, are almost identical in the MC and RP models
in the inner 100 pc.  This occurs even though the addition of momentum
feedback creates a $v_{\rm rms}$ four times larger than the MC
simulation.  This behavior is caused two complimentary processes.
First, radiation pressure creates a more diffuse medium in which the
SN explosions occur, which results in a lower cooling rate and keeps
the temperatures and, equivalently, sound speeds high.  Second, the
additional driving from radiation pressure and blastwaves that have
not stalled in the ISM keep the turbulent velocities nearly equal to
the halo circular velocity.  These physical processes lead to an
unchanged turbulent Mach in both star forming environments.

The velocity dispersion is a good measure of the general strength of
turbulent motions, but the vorticity $\bomega$ better characterises
the local strength of turbulence.  In Figure \ref{fig:vort}, we plot
two-dimensional histograms of vorticity (top panels) as a function of
radius and its growth (bottom panels), which is the right hand side of
Equation (\ref{eqn:vort}).  For the RP simulation, we include the
baroclinic term from radiation pressure in our analysis,
$(\mathbf{a}_{\rm rp} \times \nabla \rho) / \rho$.  To visualise the
fluid flow, it is helpful to consider the local rotational period is
$4\pi/|\bomega|$.  Gas with the highest vorticities exist at $100 \la
\textrm{r/pc} \la 500$.  This is primarily driven by virial shocks as
matter is accreted from the IGM and filaments.  This region is
approximately where the density increases by an order of magnitude in
the density projections of Figure \ref{fig:proj}.  The MC and RP
simulations have similar average vorticities until $r \la 50$ pc,
where the RP simulation exhibits an increase of 10, whereas the MC
simulation stabilises at $2 \times 10^{-14} \; \textrm{s}^{-1}$.
Vorticity generation is significantly higher by a factor of 100 within
this inner region with radiation pressure.  We find that
non-barotropic flows from thermal pressure are primarily responsible
for sustaining these turbulent motions.  The contribution from
vorticity stretching and radiation pressure provides $\sim$$10^{-1}$
and $\sim$$10^{-5}$ of $D\bomega/Dt$, respectively.  The very weak
driving from radiation pressure does not necessarily indicate that it
can be neglected.  Because it adds momentum to the shells, the shocks
during shell collisions are stronger and produce more vorticity
through the thermal baroclinic term.  The weak contribution from
radiation pressure is straightforward to understand.  It is strongest
when the ionisation front radius is small, and the \hii~is close to
spherical.  In the spherically symmetric case, radiation will always
be absorbed parallel to the density gradient as the optical depth
increases, and thus $\mathbf{a}_{\rm rp} \times \nabla \rho$ is small.
In a clumpy medium, this term will be larger when radiation grazes the
surface of dense clumps.  It should be noted that by neglecting
scattering we could be underestimating its magnitude, and our
calculations are conservative in quantifying the impact of radiation
pressure in the ISM of dwarf galaxies.

\subsubsection{Metal mixing}

The gas metallicity shown in Figure \ref{fig:proj} show that the SN
ejecta has expanded beyond the virial radius with outflows with
velocities up to 250 \kms, enriching most of the gas to [Z/H] $\sim$
--2.  This behaviour is further illustrated in Figure \ref{fig:phase},
where the SN ejecta thoroughly mixes with the ISM at all
density-temperature pairs in the cold and warm phases.  This greatly
differs from the lack of mixing in the MC simulation that creates a
distinct separation between dense metal-rich gas and diffuse
metal-poor gas.

Metal mixing is enhanced by a sustained turbulent medium, resulting in
a MDF (Fig. \ref{fig:mdf}) with a mean metallicity of [Z/H] = $-2.1$
and a standard deviation of 0.2 dex.  This is in excellent agreement
with the luminosity-metallicity relationship for dwarf galaxies for a
stellar system of $M_\star = 4.5 \times 10^5 \Ms$ \citep{Kirby11}.
This agreement suggests that radiation pressure plays an important
role in regulating star formation in high-redshift dwarf galaxies.

\subsubsection{Star formation regulation}

We have demonstrated that momentum input from massive stars prevents
the overcooling and over-enrichment of the gas in galaxy formation
simulations.  The additional momentum from radiation both disperses
the metal-rich SN ejecta to large radii and provides turbulent support
to the system.  The first effect reduces the metallicity in the dense
gas.  This reduces the radiative cooling from metal species and the
amount of cold, dense gas available for star formation.  The latter
effect prevents the gaseous core from catastrophically collapsing and
forming stars at a high efficiency.

Our comparison between the MC and RP simulations shows that star
formation is self-regulated with radiation pressure playing a vital
role in early galaxy formation.  To reiterate, the RP simulation
creates a galaxy with a sSFR = 37~\ssfr~and a stellar population whose
MDF ($\langle Z/Z_\odot \rangle = -2.1$) is in excellent agreement
with observations of local dwarf galaxies.

\subsection{Radiation pressure characteristics}

\begin{figure}
  \plotone{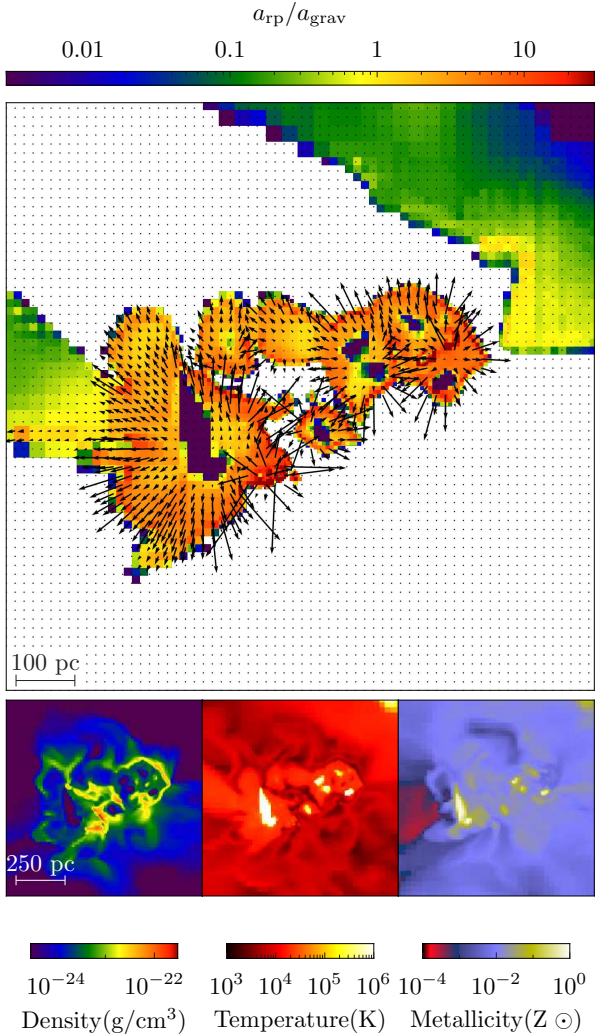}
  \caption{\label{fig:aslice} Top panel: Slices of instantaneous
    acceleration from momentum transfer from absorbed ionising
    radiation, which has been scaled to the total gravitational
    acceleration of the halo, $a_{\rm grav} = G M_{\rm vir} / r_{\rm
      vir}^2$.  Larger values indicate a greater outward acceleration
    from the responsible star particle.  The arrows denote the
    direction and magnitude of the acceleration field.  The field of
    view is 1 kpc and is cut through the halo's centre of mass.
    Bottom panels: The same field of view for gas density (left),
    temperature (middle), and metallicity (right).}
\end{figure}

\begin{figure}
  \plotone{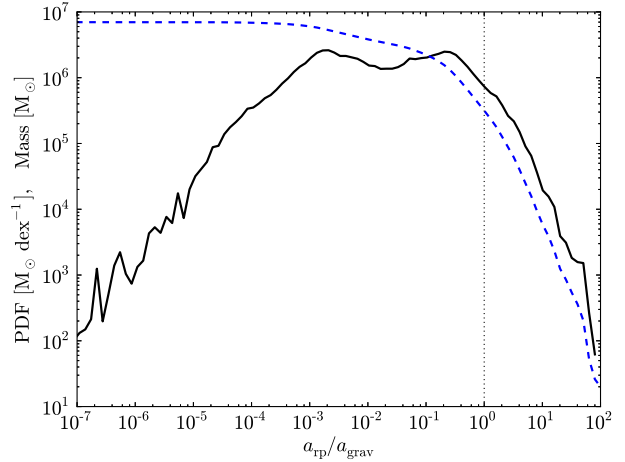}
  \caption{\label{fig:accel} Mass-weighted probability density
    distribution (solid line) of the instantaneous acceleration from
    momentum transfer from absorbed ionised radiation, which has been
    scaled to $a_{\rm grav} = G M_{\rm vir} / r_{\rm vir}^2 = 7.0
    \times 10^{-10} \; \textrm{cm} \: \textrm{s}^{-2}$, inside the virial
    radius of the most massive galaxy.  The inverse cumulative
    distribution is shown by the dashed line.  Four per cent ($\sim 3
    \times 10^5 \Ms$) of the gas is experiencing an acceleration
    greater than the total gravitational force of the halo.}
\end{figure}

Here we describe some of the properties of radiation pressure within
our largest galaxy.  Figure \ref{fig:aslice} shows two-dimensional
slices of acceleration from momentum transfer \arp~(Equation
\ref{eqn:arp}), gas density, temperature, and total metallicity
through the centre of mass of the galaxy.  The acceleration \arp~is
scaled with respect to the gravitational acceleration at the virial
radius, $a_{\rm grav} = GM_{\rm vir}/r_{\rm vir}^2 = 7.0 \times
10^{-10} \; \textrm{cm} \: \textrm{s}^{-2}$.  The time to
accelerate the gas above the escape velocity $v_{\rm esc} = (2GM_{\rm
  vir}/r_{\rm vir})^{1/2}$ is
\begin{equation}
t_{\rm esc} = v_{\rm esc}/a_{\rm grav} = [50 \Omega_m(z) H^2(z)]^{-1/2}
\end{equation}
which is independent of the halo.  Here $\Omega_m(z)$ and $H(z)$ are
the matter density and Hubble constant at redshift $z$.  In a matter
dominated universe ($z \gg 1$) for the adopted \lcdm~cosmology, this
timescale is $120 [(1+z)/10]^{-3/2}$ Myr, which is possible with
sustained star formation in a galaxy.  However, this should be only be
used to gauge the instantaneous impact of the momentum transfer
because the acceleration will decrease as the shell travels farther
from the source.

The density plot shows that the ISM is fractal and porous in nature
with the \hii~regions, creating dense shells on their surfaces.  The
acceleration \arp~exceeds 10 per cent of \agrav~in most of the
\hii~region except for the interiors of the SN remnants, where the
neutral hydrogen fraction is nearly zero.  It exceeds \agrav~in the
dense, neutral shells surrounding the \hii~regions.  Figure
\ref{fig:accel} shows the probability and cumulative distribution
function of \arp/\agrav~inside of this halo.  There is a plateau in
the distribution between $2 \times 10^{-3}$ and $0.3$~at $2 \times
10^6$ \Ms~per dex.  Radiation pressure is accelerating about 4 per
cent ($\sim 3 \times 10^5 \Ms$) of the halo gas greater than \agrav.

The slices of temperature and total metallicity in Figure
\ref{fig:aslice} show the hot phase of the ISM, which have $T > 10^6$
K and approximately solar metallicity.  Our stellar feedback
prescription injects SN thermal energy and metals after 4 Myr of main
sequence radiative feedback.  Consequently, the SN remnants are
initially contained within the rarefied medium of the interior of
their associated \hii~regions and then expand into the ISM, and in
some cases, they drive metal-rich outflows into the IGM.

\section{Discussion}
\label{sec:discuss}

Stellar feedback comes in various forms.  First, photo-ionisation and
photo-heating from massive stars create \hii~regions that provide
pressure support to the warm phase of the ISM.  An extreme case of
photo-heating feedback exists in haloes with $M \la 10^7 \Ms$ that host
Pop III stars, where the $\sim$30 \kms~shock launched by the
\hii~region can expel over 90 per cent of the gas from the halo
\citep{Whalen04, Kitayama04, Abel07}.  Second, SNe feedback injects
thermal energy and heavy elements into the ISM, creating the hot
bubbles and blastwaves.  They can self-enrich their environment
\citep[e.g.][]{Wise12_Galaxy}, disrupt its star-forming cloud
\citep[e.g.][]{Draine91, Whalen08_SN}, create realistic bulgeless disc
galaxies, and possibly launch large-scale outflows \citep{Governato10,
  Brook12}.  Both processes together can play an important role in
regulating star formation and altering gas and stellar metallicities,
galactic morphologies, gas fractions, and the structure of the ISM.

Another feedback process is the momentum transfer from ionising
radiation from massive stars, i.e. radiation pressure.  It has been
suggested that radiation pressure can regulate star formation and
launch galactic outflows in large galaxies \citep{Haehnelt95,
  Murray05, Murray11, Hopkins11_RP}.  In this paper, we have
investigated the role of momentum deposition from ionising radiation
in the formation of high-redshift dwarf galaxies.  We have found that
the addition of radiation pressure regulates star formation through
additional turbulent support from shell collisions and allowing for
SNe to drive large-scale outflows from the galaxy, ejecting metal-rich
gas in the process.  These effects alleviate the galaxy overcooling
problem seen in various simulations of galaxy formation.

By including radiation pressure, our simulation creates realistic
high-redshift dwarf galaxies that matches the local
luminosity-metallicity relation.  Although this is not a direct
comparison between high-redshift and local dwarf galaxies, the
metallicities of the stellar population is a good gauge of a
reasonable ISM chemo-thermal state.  If we assume that a fraction of
high-redshift galaxies survive until the present-day \citep{Gnedin06,
  Bovill11b}, then the bounds of MDFs should not exceed the ones
observed in local dwarf galaxies, which typically have most of their
stars within the range --3 $<$ [Fe/H] $<$ --1 \citep{Kirby11}.  For
instance in our MC simulation, the stars have a median metallicity of
$\sim$0.3 \zsun, which will still exist in its MDF if we were to
passively evolve them to $z=0$.  On the other hand, this $M = 2 \times
10^8 \Ms$ galaxy is most likely to be incorporated into a galactic
bulge, based on its halo mass accretion history \citep{Wechsler02}.
However, this result of overcooling is generic to galaxies reaching $T
= 10^4$ K and should occur in dwarf galaxies that form later.  Another
possible way to avoid this restriction is that these stars form from a
top-heavy IMF.  If all of the stars form with $M \ga 0.8 \Ms$, then
they will not survive until the present-day, and the MDFs from
high-redshift dwarfs cannot be constrained by the observed MDFs in
local dwarf galaxies.

\subsection{Feedback prescriptions}

Implementations of stellar feedback vary widely between different
works.  Here we discuss these variations, and their respective
advantages and disadvantages.  As galaxy formation simulations reach
higher resolution, individual star forming clouds are being resolved.
For example, \citet{Governato10} found that a bulgeless disc galaxy
forms instead of a bulge-dominated galaxy when the star formation
density threshold was increased by a factor of 1000 to 100 \cubecm.
This occurred because the simulation allowed the gas to cool and clump
into smaller associations instead of a large gas reservoir at the
galaxy centre.  When simulations resolve these clouds, more
small-scale physical processes need to be accurately modelled, and the
less previous ``subgrid'' models are applicable because of their
original assumptions of star formation on the galactic scale.

As discussed before, there are four main components to stellar
feedback: photo-ionisation, photo-heating, momentum transfer from
radiation, and SN explosions, in addition to stellar winds and
planetary nebulae.  The temporal separation of the radiative and SNe
feedback is \textit{crucial}.  In the case where both mechanisms have
the same time-dependency, the injection of thermal energy and heavy
elements are initially dumped into a dense and cold medium, unaffected
by stellar radiation.  If the SNe feedback is delayed for some amount
of time, in our case, the minimum lifetime of an O-star of 4 Myr, then
an \hii~region is created surrounding the star(s), which lowers the
ambient density and increases the temperature to $\sim$10$^4$ K.  An
optically-thick shell forms around the \hii~region, and radiation
pressure accelerates the shell in addition to pressure forces from the
\hii~region \citep{Spitzer78}.  After a few Myr, the massive stars
start to explode in this warm and diffuse medium, instead of a cold
and dense medium.  Because cooling rates are $\propto \rho^2$, the SN
remnants are less likely to overcool in an implementation that delays
SNe feedback from radiative feedback.

There have been a few types of prescriptions of SNe and momentum
injection to create galactic winds and to regulate star formation.
For the most part, they have an unphysical basis but replicate the
observed properties of outflows and cannot study the origin of the
winds.  

In galaxy simulations, it has been historically difficult to resolve
the multi-phase ISM.  In such a simulation, SNe thermal energy is
injected into a large region, usually at its resolution limit.  Per
timestep, the total thermal energy is only sufficient to heat the gas
to $T \sim 10^5 - 10^6$ K, where the cooling rates are maximal.  If
the blastwave were resolved to parsec-scales, then the temperature in
the Sedov-Taylor solution should be $T \sim 10^8$ K, where the cooling
rates are orders of magnitude lower.  One approach is to neglect
radiative cooling within the blastwave, allowing it to adiabatically
expand \citep[e.g.][]{Thacker00, Governato07, Guedes11}.  Using this
method results in realistic late-type galaxy morphologies, but it can
deviate from the observed luminosity functions because they do not
suppress star formation enough, particularly in low mass galaxies, and
do not drive winds at sufficiently high velocities and mass-loading
factors \citep[e.g.][]{Guo10}.  We note that another approach is to
sporadically inject thermal energy only after enough SNe thermal
feedback has accumulated to heat the gas to $T \ga 10^7$~K, where the
cooling times are high.  Here ejecta does not overcool and follows the
Sedov-Taylor solution \citep{Dalla12}.

Several groups \citep{Springel03b, Oppenheimer08, DallaVecchia08,
  Sales10} have modelled momentum transfer from radiation by adding
momentum, so-called ``kicking'', to nearby particles to a velocity
that is comparable to the circular velocity $V_c$ of the halo.
Similar prescriptions exist for kinetic feedback from SN explosions
\citep{DallaVecchia08}.  The dependence on the choice of this kick
velocity, a constant value, a constant multiple of $V_c$, or halo mass
dependent velocity, has been shown to replicate the galaxy luminosity
function and mass-metallicity relation \citep{Finlator08}.  The kicked
particles are decoupled from the hydrodynamics until they escape the
virial radius.  The authors acknowledge and warn that this
implementation is phenomenological, and it is successful in
quantifying the importance of momentum-driven winds in MW-like
galaxies.  \citet{Hopkins11_RP} used a more physical model for
momentum loading that does not decouple the winds from the
hydrodynamics and determines the momentum transfer directly from the
stellar luminosity.  The momentum transfer occurs within $\la$10 pc,
where they explore two schemes.  Their stochastic model randomly kicks
these nearby particles to the local escape velocity of the
star-forming clump.  Their continuous model adds the momentum $\Delta
p = (1 + \eta_p \tau_d) L / c$ at each timestep, where $\eta_p \sim 1$
and can vary to account to uncertainties.  They find that both models
produce similar results.  However, they do not solve the radiative
transfer equation, so they must assume that the radiation is uniform
within each \hii~region.  Radiation pressure from a non-uniform
distribution of massive stars may result in some cancellation of the
pressure forces in the centre of the association, while still
providing an overall outward force \citep{Socrates08}.  In contrast,
\citet{Krumholz12_RP} showed that the assumption of a uniform
radiation field could lead to an overestimate the momentum transfer.
As ionisation front instabilities grow, optically-thin gaps are
created in the shock front, allowing for most of the flux to escape in
these gaps and lowering the average momentum transfer per emitted
photon.

Ultimately the location where momentum transfer occurs depends where
the radiation is absorbed.  To model this requires solving the
radiative transfer equation.  None of these models with the exception
of \citeauthor{Krumholz12_RP} utilises such a solver, but there are
several simulation codes with coupled radiation transport.  Momentum
transfer from ionising radiation should be straightforward to
implement in these codes because the absorption coefficient is already
calculated and that absorbed photon momentum should be transferred to
the absorbing medium, which is how \moray~simulates momentum loading.

\subsection{Radiation trapping from dust and Lyman $\alpha$}
\label{sec:discuss_trap}

\begin{figure}
  \plotone{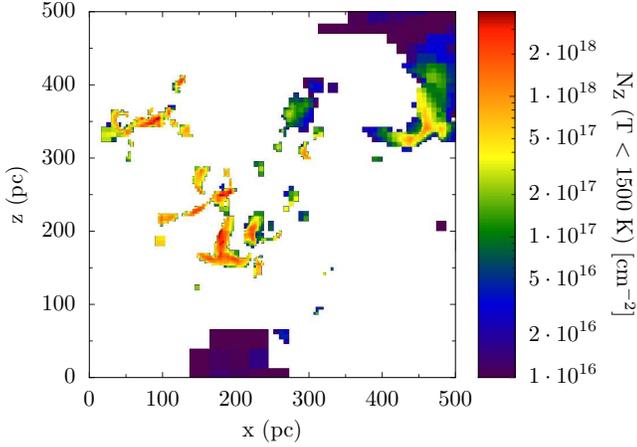}
  \caption{\label{fig:zcol} Column density of metals with $T < 1500$
    K, where dust would not be sublimated and can contribute to the
    opacity and enhance momentum transfer from stellar radiation.  The
    figure is centred on the galaxy centre of mass.}
\end{figure}

In this paper, we have neglected momentum transfer from dust
absorption, which then re-radiates in the IR and absorbed again many
times.  Thus our results represent a lower limit of star formation
regulation from radiation pressure.  The effective cross section of
dust in the Milky Way is $\sigma_{d, \rm {MW}} \simeq 10^{-23}$
cm$^{2}$ per hydrogen atom in the near-IR \citep[e.g.][]{Draine03}, or
equivalently, an opacity $\kappa_{d, \rm{MW}} \simeq 5$ cm$^{2}$
g$^{-1}$ in terms of total gas mass.  It should be noted that
Rosseland dust opacity is temperature dependent, scaling as $T^2$
below 150 K and can vary by factors of a few at $150 \textrm{K} < T <
1500 \textrm{K}$.  Above 1500 K, even the most stable dust species are
sublimated and the Rosseland opacity drops by a few orders of
magnitude \citep{Semenov03}.  If we assume the dust extinction
properties do not change with metallicity or redshift and scale the MW
values with metallicity, the cross section per hydrogen atom scales
with metallicity, giving $\kappa_d = (Z/\mathrm{Z}_\odot) \kappa_{d,
  \rm{MW}}$.  The optical depth to dust absorption is
\begin{equation}
  \tau_d = \int \rho \; \kappa_d \; dl = \kappa_{d, \rm{MW}} \;
  \frac{\mu m_{\rm H} N_Z}{\textrm{Z}_\odot},
\end{equation}
where $N_Z$ is the column density of metals, and $\mu$ and $m_H$ are
the mean molecular weight and mass of a hydrogen atom.  We take $\zsun
= 0.023$ \citep{Zsun}.  Figure \ref{fig:zcol} shows the metal column
density $N_Z$ of cold ($T < 1500$ K) gas in the inner 500 pc of the
galaxy.  Dust in gas $T > 1500$ K is destroyed and is not considered
in the plot.  Using $\kappa_{d, \rm {MW}} = 5 \; \textrm{cm}^2 \:
\textrm{g}^{-1}$, the column density range $N_Z = (10^{16}, 10^{18.6})
\; \textrm{cm}^{-2}$ corresponds to a optical depth range $\tau_d =
(10^{-5.3}, 10^{-2.7})$.  In this case, the dust absorption should
have little effect on radiation momentum transfer in dwarf galaxies
and only becomes significant in MW-like galaxies.  However, dust
opacities may reach several hundred cm$^2$ g$^{-1}$ in local and
distant starburst galaxies \citep[e.g.][]{Lehnert96, Sanders96,
  Calzetti01}; however, see \citet{Krumholz12_RP} for a
counter-argument.  In this maximal case, the optical depth $\tau_d$ is
of order unity and could boost the momentum transfer by a factor of a
$\sim$few if the dust were to couple with the gas
\citep[e.g.][]{Murray05}.

In these low metallicity or dust-depleted cases, Lyman~$\alpha$
radiation can be trapped through resonant interactions, producing a
possible significant source of radiation pressure \citep{Henney98}.
This process becomes insignificant when dust opacities reach unity,
where dust reprocesses UV starlight into IR.  Lyman~$\alpha$ radiation
accumulates in the resonances, and eventually its pressure saturates,
enhancing momentum transfer by a factor of
\begin{equation}
  f_{\rm trap, Ly \alpha} \approx 0.06 \
  \left( \frac{\sigma_{\rm d}}{\sigma_{\rm d, MW}} \right)^{-1} \
  \left( \frac{r}{R_{\rm eq}} \right)^{1/2},
\end{equation}
where $R_{\rm eq}$ is the radius at where the gas pressure equals
radiation pressure \citep{Krumholz09_RP}.  Thus we could be further
underestimating radiation pressure as Lyman~$\alpha$ trapping should
become important below $\sim 0.05 Z_\odot$, given similar dust
properties as the MW.

\section{Conclusions}
\label{sec:conclusions}

We present results from three cosmological radiation hydrodynamics
simulations that demonstrate the role of radiation pressure in
affecting star formation in dwarf galaxies.  We progressively add more
physics to our simulations to understand the impact of each process.
Our reference model considers primordial chemistry and Pop II and III
star formation and feedback with a transition to a Salpeter IMF at
$10^{-4} \zsun$.  In the next model, we add radiative cooling from
metals.  In our final and most realistic model, we add an
\hh-dissociating radiation background and momentum input from stellar
radiation.  Our treatment of momentum transfer from ionising radiation
is accurately calculated from the optical depth given by our radiation
transport module \moray~down to our resolution limit of 1 comoving
parsec.  To our knowledge, this is the first cosmological galaxy
simulation that has included the effects of radiation pressure that is
computed from the radiative transfer equation.

The additional radiative cooling from metal species induced a runaway
star formation event with a sSFR = 150 \ssfr~and SFR = $9.9 \times
10^{-2}$ \hsfr~in our most massive galaxy with $M_{\rm DM} = 2.0
\times 10^8 \Ms$ at $z = 8.1$.  This strong burst creates enough
ionising photons to reionise the entire 1 Mpc$^3$ comoving domain;
however, it creates a stellar population with a mean metallicity of
0.3~\zsun~that does not follow the mass-metallicity relationship of
dwarf galaxies.  The latter fact indicates that the galaxy is
overcooling and overproducing stars.  Historically, this was avoided
by using higher resolution and SNe feedback, but our simulations have
both of these attributes.  We find that \textit{radiation pressure
  plays a key role in regulating star formation} in high-redshift
dwarf galaxies.  The highlights of our work on the effects of
radiation pressure during the formation of a dwarf galaxy are as
follows.

\begin{enumerate}
\item Radiation pressure alone can reduce the SFR by a factor of
  $\sim$5 to $1.8 \times 10^{-2}$ \hsfr, and a similar decrease is
  seen in the sSFR to 37 \ssfr.  About 3 per cent of the accreted gas
  form stars in the buildup of the stellar population.
\item The collision of dense shells that are driven by radiation
  pressure creates a turbulent medium with an rms velocity equivalent
  to the circular velocity of the host DM halo.  This added turbulence
  mixes the SNe ejecta thoroughly with the entire ISM.  The turbulent
  pressure and reduced metallicity prevents overcooling and the
  subsequent catastrophic collapse.
\item Radiation pressure in dwarf galaxies is not sufficient to launch
  outflows, but it provides an additional impulse to drive dense gas
  away from massive stars and drive interstellar turbulence.  SNe
  energy input is still the main mechanism to drive outflows in
  low-mass galaxies.
\item Star formation regulation by radiation pressure creates a
  stellar population with a mean metallicity of $10^{-2.1} \zsun$ with
  a normal metallicity distribution with standard deviation of 0.2
  dex, agreeing with the mass-metallicity relationship of local dwarf
  galaxies.
\item We show that the majority of the momentum transfer occurs in
  dense shells at the edges of \hii~regions, i.e. it is not a local
  process.  Furthermore, the timing of numerical radiative and SNe
  feedback processes is critical in capturing their effects.  If the
  SNe feedback were to be released immediately into the ISM, densities
  will be artificially high and the SNe input will radiate too much of
  its energy.  In nature, \hii~region dynamics will evacuate the
  immediate volume surrounding massive stars of dense and cold gas,
  and SN explosions occur in a warmer, more diffuse ISM.  Our
  simulations capture this temporal sequence, and the SN blastwaves do
  not overcool and are able to launch outflows from the galaxy.
\item The column density of metals in $10^8 - 10^9 \Ms$ dwarf galaxies
  reach a maximum value of $\sim 10^{19}$ cm$^{-2}$.  This may imply
  that that momentum transfer via dust grains is not important in
  low-mass early galaxies.
\end{enumerate}

We have shown that radiation pressure regulates star formation in
dwarf galaxies in addition to photo-heating and SNe feedback.  Its
inclusion in galaxy formation simulations is key in forming realistic
stellar populations and avoiding the overcooling problem also in
high-resolution simulations that capture star forming regions with
many computational elements.  The impact of radiation pressure should
continue to be important in larger galaxies, and we plan to apply our
numerical methods to simulating such galaxies that are currently
observed in the \textit{Hubble Ultra Deep Field}.

\section*{Acknowledgments}

We thank Mark Krumholz and an anonymous referee for helpful comments
on this manuscript.  When this work started, JHW was supported by NASA
through Hubble Fellowship grant \#120-6370 awarded by the Space
Telescope Science Institute, which is operated by the Association of
Universities for Research in Astronomy, Inc., for NASA, under contract
NAS 5-26555.  MJT~acknowledges support by the NSF CI TraCS fellowship
award OCI-1048505.  Computational resources were provided by NASA/NCCS
award SMD-11-2258 and a director's discretionary allocation on SDSC
Trestles.  This work was partially supported by NASA ATFP grant
NNX08AH26G, NSF AST-0807312, and NSF AST-1109243. This research has
made use of NASA’s Astrophysics Data System Bibliographic Services.
The majority of the analysis and plots were done with \texttt{yt}
\citep{yt_full_paper}.

\bibliography{ms}
\bsp
\label{lastpage}

\end{document}

%% file: macros.tex
\newcommand{\fesc}{\ifmmode{f_{\rm esc}}\else{$f_{\rm esc}$}\fi}
\newcommand{\fescs}{\ifmmode{f_{\rm esc}^\star}\else{$f_{\rm esc}^\star$}\fi}
\newcommand{\kms}{\ifmmode{{\;\rm km~s^{-1}}}\else{km~s$^{-1}$}\fi}
\newcommand{\fgas}{\ifmmode{{f_{\rm gas}}}\else{$f_{\rm gas}$}\fi}
\newcommand{\cubecm}{\ifmmode{{\rm cm^{-3}}}\else{cm$^{-3}$}\fi}
\newcommand{\ztwo}{\ifmmode{{\rm [Z_2/H]}}\else{[Z$_2$/H]}\fi}
\newcommand{\zthree}{\ifmmode{{\rm [Z_3/H]}}\else{[Z$_3$/H]}\fi}
\newcommand{\lsim}{\lower0.3em\hbox{$\,\buildrel <\over\sim\,$}}
\newcommand{\gsim}{\lower0.3em\hbox{$\,\buildrel >\over\sim\,$}}

\newcommand{\lcdm}{$\Lambda$CDM}

\newcommand{\emis}{erg s$^{-1}$ cm$^{-2}$ Hz$^{-1}$ sr$^{-1}$}

\newcommand{\hsfr}{M$_\odot$ yr$^{-1}$}
\newcommand{\ssfr}{Gyr$^{-1}$}
\newcommand{\arp}{$a_{\rm rp}$}
\newcommand{\agrav}{$a_{\rm grav}$}
\newcommand{\eavg}{\ifmmode{\langle E_\gamma \rangle}\else{$\langle E_\gamma \rangle$}\fi}

\newcommand{\enzo}{{\sc enzo}}
\newcommand{\moray}{{\sc enzo+moray}}
\newcommand{\Ms}{\ifmmode{M_\odot}\else{$M_\odot$}\fi}
\newcommand{\vrms}{\ifmmode{v_{\rm rms}}\else{$v_{\rm rms}$}\fi}

\newcommand{\hh}{H$_2$}

\newcommand{\tvir}{\ifmmode{T_{\rm{vir}}}\else{$T_{\rm{vir}}$}\fi}
\newcommand{\mvir}{\ifmmode{M_{\rm{vir}}}\else{$M_{\rm{vir}}$}\fi}
\newcommand{\rvir}{\ifmmode{r_{\rm{vir}}}\else{$r_{\rm{vir}}$}\fi}

\newcommand{\jj}{\ifmmode{J_{21}}\else{$J_{21}$}\fi}
\newcommand{\flw}{\ifmmode{F_{LW}}\else{$F_{LW}$}\fi}
\newcommand{\kph}{\ifmmode{k_{\rm ph}}\else{$k_{\rm ph}$}\fi}

\newcommand{\msun}{{\rm\,M_\odot}} 
\newcommand{\lsun}{{\rm\,L_\odot}}
\newcommand{\zsun}{\ifmmode{\rm\,Z_\odot}\else{$\rm\,Z_\odot$}\fi}

\newcommand\tento[1]{$10^{#1}$}

\newcommand{\hi}{H {\sc i}}
\newcommand{\hii}{H {\sc ii}}
\newcommand{\hei}{He {\sc i}}
\newcommand{\heii}{He {\sc ii}}
\newcommand{\heiii}{He {\sc iii}}

\def\eps@scaling{1.0}%
\newcommand\epsscale[1]{\gdef\eps@scaling{#1}}%
\newcommand\plotone[1]{%
 \centering 
 \leavevmode 
 \includegraphics[width={\eps@scaling\columnwidth}]{#1}%
}%
\newcommand\plottwo[2]{%
 \centering 
 \includegraphics[width={\eps@scaling\columnwidth}]{#1}%
 \hfil 
 \includegraphics[width={\eps@scaling\columnwidth}]{#2}%
}%